\documentclass[12pt,preprint]{aastex}

\newcommand{\mh}{H$_{2}$~}
\newcommand{\mhs}{H$_{2}$}

\newcommand{\msolars}{M$_{\odot}$}
\newcommand{\msolar}{M$_{\odot}$~}

\newcommand{\um}{$\mu$m~}
\newcommand{\ums}{$\mu$m}

\def\h0{{\rm H_0}}

\def\pm{^+_-}
\def\ciil{[C\kern.2em{\sc ii}] 158 $\mu$m }
\def\cii{[C\kern.2em{\sc ii}]}
\def\nii{[Ne\kern.2em{\sc ii}]}
\def\oii{[O\kern.2em{\sc ii}] }

\def\oiii{[O\kern.2em{\sc iii}] }

\def\oiiil{[O\kern.2em{\sc iii}] 88.356 $\mu$m }

\def\oiiif{[O\kern.2em{\sc iii}] \kern.2em{3p2-3p1}}

\def\oiiie{[O\kern.2em{\sc iii}] \kern.2em{3p1-3p0}}

\def\oiv{[O\kern.2em{\sc iv}] }

\def\siiil{[S\kern.2em{\sc iii}] 18.7130 $\mu$m}

\def\s33l{[S\kern.2em{\sc iii}] 33.4810 $\mu$m}

\def\sivl{[S\kern.2em{\sc iv}]  10.5105  $\mu$m}

\def\s2l{[Si\kern.2em{\sc ii}]  34.8152  $\mu$m}

\def\neiil{[Ne\kern.2em{\sc ii}]  12.8135  $\mu$m}
\def\neiiil{[Ne\kern.2em{\sc iii}]  15.5551  $\mu$m}

\def\siii{[S\kern.2em{\sc iii}] }

\def\siv{[S\kern.2em{\sc iv}] }

\def\si2{[Si\kern.2em{\sc ii}]}

\def\neii{[Ne\kern.2em{\sc ii}]  }
\def\neiii{[Ne\kern.2em{\sc iii}] }

\shorttitle{Molecular Hydrogen}
\shortauthors{Higdon et al.}

\begin{document}

\title{A Spitzer\footnotemark [1] ~Infrared Spectrograph\footnotemark [2] ~Survey of Warm Molecular Hydrogen in Ultra-luminous Infrared Galaxies}

\footnotetext [1] {Based on
       observations obtained with the Spitzer Space Telescope, which
       is operated by JPL, California Institute of Technology for the
       National Aeronautics and Space Administration.}

\footnotetext [2] {The IRS was a
       collaborative venture between Cornell University and Ball
       Aerospace Corporation funded by NASA through the Jet Propulsion
       Laboratory and the Ames Research Center.}

\author{S. J. U. Higdon\altaffilmark{3}, L. Armus\altaffilmark{4},
J. L. Higdon\altaffilmark{3},  B. T. Soifer\altaffilmark{4}
\& H. W. W. Spoon\altaffilmark{3}}

% T. Herter\altaffilmark{1}, V. Charmandaris\altaffilmark{3},

\altaffiltext{3}{Astronomy Department, Cornell University, Ithaca, NY 14853}
%\altaffiltext{4}{Chercheur Associ\'e, Observatoire de Paris, F-75014, Paris, France}
\altaffiltext{4}{Spitzer Science Center, California Institute of Technology, 220-6, Pasadena, CA 91125}
%\altaffiltext{3}{University of Crete}

\begin{abstract}
  We have conducted a survey of Ultra-luminous Infrared Galaxies
  (ULIRGs) with the Infrared Spectrograph on the Spitzer Space
  Telescope, obtaining spectra from 5.0$-$38.5~\um for 77 sources with
  0.02$<$z $<$0.93. Observations of the pure rotational~\mh lines
  S(3)~9.67~\ums, S(2)~12.28~\um, and S(1)~17.04~\um are used to
  derive the temperature and mass of the warm molecular gas. We detect
  \mh in 77\% of the sample, and all ULIRGs with F$_{60 \mu m}>2$~Jy.
  The average warm molecular gas mass is
  $\sim$2~$\times~10^8$~\msolars. High extinction, inferred from the
  9.7~$\mu$m silicate absorption depth, is not observed along the line
  of site to the molecular gas. The derived \mh mass does not depend
  on F$_{25 \mu m}/$F$_{60 \mu m}$, which has been used to infer
  either starburst or AGN dominance. Similarly, the molecular mass
  does not scale with the 25~\um or 60~\um luminosities. In general,
  the \mh emission is consistent with an origin in photo-dissociation
  regions associated with star formation. We detect the S(0)~28.22 \um
  emission line in a few ULIRGs. Including this line in the model fits
  tends to lower the temperature by $\sim$50-100~K, resulting in a
  significant increase in the gas mass. The presence of a cooler
  component cannot be ruled out in the remainder of our sample, for
  which we do not detect the S(0)~line. The measured
  S(7)~5.51~\um~line fluxes in six ULIRGs implies
  $\sim$3$\times$~$10^6$~\msolar of hot ($\sim$~1400~K)~H$_{\rm 2}$.
  The warm gas mass is typically less than one~percent of the cold gas
  mass derived from $^{12}$CO observations.
\end{abstract}


\section{Introduction}

Ultra-luminous infrared galaxies (ULIRGs) are the most luminous
objects in the local universe. These objects appear to be mergers of
dusty, gas-rich disk galaxies (e.g., Sanders \& Mirabel 1996; Moorwood
1996). A direct result of this dusty and large optical depth
environment is that they radiate 90\% or more of their total energy in
the far-infrared, typically with L$_{8-1000 \mu m}$ $\ge$ 10$^{12}$
L$_{\odot}$.  The bolometric luminosities and space densities of
ULIRGs in the local universe (i.e., z $\le$ 0.1) are similar to those
of QSOs (Soifer et al.  1987; Sanders \& Mirabel 1996). It was
proposed by Sanders et al. (1988) that most ULIRGs are powered by
dust-enshrouded QSO's in the late phases of a merger. A classic
example of this is Mrk 231, where the detection of broad ($\sim$
10,000 km s$^{-1}$) hydrogen emission lines (Arakelian et al. 1971;
Adams \& Weedman 1972) and radio jets (Lonsdale et al. 2003) provides
direct evidence for the central active galactic nucleus (AGN). The
final state of a ULIRG may be a large elliptical galaxy with a central
massive quiescent black hole (Kormendy \& Sanders 1992).  The ratio of
the 25 \um to 60 \um flux densities, as measured with the Infrared
Astronomical Satellite (IRAS, Neugebauer et al. 1984), has been used
to classify ULIRGs. This ratio reflects the globally averaged dust
temperature of the galaxy, which is typically 50 - 100 K.  The ULIRGs
are classified as either ``warm'' (F$_{25 \mu m}$/F$_{60 \mu m}$ $\ge$
0.2) and possibly AGN dominated, or ``cold'' (F$_{25 \mu m}$/F$_{60
  \mu m}$ $<$ 0.2), with a possible dominant contribution to the
overall luminosity from massive star formation. Although luminous
mergers are rare in the local universe, it has been proposed that
ULIRGs may make a significant contribution to the star formation
density at z $\ge$ 1 (e.g., Elbaz \& Cesarsky 2003).

Molecular hydrogen is the primary component of the dense gas in
galaxies and the most abundant molecule in the Universe. Stars are not
only formed from molecular clouds, but the \mh acts as a coolant
enabling the star formation to proceed. \mh may also act as a coolant
facilitating the accretion of material onto a central AGN. Prior to
space borne missions large quantities of molecular hydrogen in the
central regions of ULIRGs has been inferred from detections of
rotational transitions of $^{12}$CO (e.g., Sanders, Scoville \& Soifer 1991; 
and references for this sample in Table 4.).  A much smaller, hot 
($\sim$2000 K) gas component can be observed directly in the near-infrared
through ro-vibrational \mh lines (e.g., Van der Werf et al. 1993).

In this paper, we have used observations with the Spitzer Space
Telescope (Werner et al. 2004) to directly probe the warm ($\ga$100 K)
molecular hydrogen component through pure rotational transitions,
where the upper levels are populated via UV pumping, formation of \mh
in excited states, or collisional excitation.  Four heating mechanisms
have been proposed for the gas: (1) grain photoelectric heating in
photo-dissociation regions (PDRs), (2) shock heating from outflows,
SNRs and large scale streaming motions in spiral arms and bars, (3)
X-ray heating of gas from AGN, SNRs and cooling flows, and (4)
cosmic-ray heating.  \mh has two independent states: ortho-\mhs, where
the nuclear spins are parallel, and para-\mhs, where the spins are
anti-parallel. The rotational states have even spatial symmetry in the
para-\mh and odd spatial symmetry in the ortho-\mhs. An ortho-to-para
ratio of 3 is found for gas of typical temperature T $\sim$ 300 K. The
proton exchange mechanism is a pathway to set up and maintain these
LTE ratios. \mh is highly symmetric and has no permanent dipole
moment.  All the rotational transitions within the electronic ground
state are quadrupolar, with correspondingly low spontaneous Einstein A
coefficients.  This makes the emission lines very weak, and until the
advent of Spitzer, difficult to observe in all but the brightest
galaxies. Valentijn et al. (1996) published the first detection of the
\mh S(0) 28.22 \um transition in the galaxy NGC~6946, using the Short
Wavelength Spectrometer (SWS) on the Infrared Space Observatory (ISO).
Subsequently, Rigopoulou et al.  (2002) published a survey of \mh in 12
starburst and 9 Seyfert galaxies using ISO. \mh has been detected in
only two ULIRGs, Arp 220 (Sturm et al. 1996) and NGC 6240 (Lutz et al.
2003). A review of the ISO \mh observations is given in Habart et al.
(2005).

In order to investigate the properties for a statistically representative
sample of the local ULIRG population, we have selected 110 ULIRGs having 
with redshifts between 0.02 and 0.93 for observation with Spitzer's
Infrared Spectrograph (IRS, Houck et al. 2004), as part of the IRS GTO
program. These sources are chosen from the complete Bright Galaxy
Sample (Soifer et al.  1987), the 1 Jy (Kim \& Sanders 1998) and 2 Jy
(Strauss et al. 1992) samples, and the FIRST/IRAS radio-far-infrared
sample of Stanford et al. (2000). Armus et al. (2004) published the
first results from this survey for Mrk~1014, Mrk~463, and UGC~5101. Two 
other papers discussing NGC~6240 and the ten ULIRGs in the IRAS bright galaxy
sample (Soifer et al. 1987) are in press (Armus et al.  2006a \&
2006b, respectively). In this paper we use the unprecedented sensitivity
of the IRS to assess the warm \mh component in a sub-sample of 77 ULIRGs.
All sources were detected at 60 \um with IRAS. Forty-eight of the ULIRGs
in our sample are classified as ``cold'' (i.e., F$_{25 \mu m}$/F$_{60 \mu m}$
$<$ 0.2) and likely starburst dominated. The remaining twenty-nine ULIRGs
are ``warm'' (i.e., F$_{25 \mu m}$/F$_{60 \mu m}$ $\ge$ 0.2), and are probably
powered by an obscured active nucleus. The basic properties of the sample
are listed in Table 1.

The Spitzer IRS wavelength coverage encompasses the pure rotational
S(0) 28.22 \um through S(7) 5.51 \um transitions. The relative line
strengths allow the determination of the temperature and mass of the
warm \mh gas. In \S 2 we detail the observations and data reduction,
and in \S 3 we present the results. In general the analysis of the
warm molecular gas is approximated with a single temperature model.
However, we detected the S(0) line in three ULIRGs, which can lower
the derived warm gas temperature.  Additionally, in some systems we
detect the S(7) line, and we use this to model a hot gas component. In
the discussion section (\S 4) we investigate the warm gas mass as a
function of the average global dust temperature. We also assess the
warm gas fraction with respect to the cold gas mass estimates
available in the literature.  Our conclusions are presented in \S 5.
Throughout the paper we adopt a flat $\Lambda$-dominated Universe
(H$_{o}$ = 71 km s$^{-1}$Mpc$^{-1}$, $\Omega_M$ = 0.27,
$\Omega_{\Lambda}$ = 0.73, and $\Omega_{k}$ = 0).

\section{Observations And Data Reduction}

The data presented here were obtained using the IRS. In low resolution
mode (IRS-LORES) there are two spectrometers with a resolving power
64 $\le$ $\frac{\lambda}{\Delta\lambda}$ $\le$ 128.  Short-low
(IRS-SL) operates between 5.2 $\mu$m and 7.7 \um in short-low 2
(IRS-SL2) and 7.4 - 14.5 $\mu$m in short-low 1 (IRS-SL1).  Long-low
(IRS-LL) gives coverage from 14.0 - 21.3 \um in long-low 2 (IRS-LL2)
and 19.5 - 38.0 $\mu$m in long-low 1 (IRS-LL1). The two high
resolution spectrometers (IRS-HIRES) have a resolving power,
$\frac{\lambda}{\Delta\lambda} \sim 600$.  Short-high (IRS-SH)
encompasses the range 9.9 - 19.6 \um and Long-high (IRS-LH) spans 18.7
- 37.2 \ums.  The observations were made in the IRS Staring Mode
Astronomical Observing Template (AOT). A high-accuracy blue peak-up,
using either a star from the 2MASS catalog (Cutri et al.  2003) or
the source itself, was executed in order to accurately place our
targets on the IRS slits. The observation log is given in Table 2. The
IRS is fully described in Houck et al. (2004) and the observing mode
details are presented in chapter 7 of the Spitzer Observers Manual
(SOM)\footnotemark[1] 
\footnotetext[1]{\url{http://ssc.spitzer.caltech.edu/documents/som}}.

The spectral data were processed as far as the un-flatfielded two
dimensional image using the standard IRS S11 pipeline (see the SOM).
The spectra were then extracted and sky subtracted using the SMART
analysis package (Higdon et al. 2004). To maximize the signal-to-noise
in the final spectra the data were extracted using a column whose
width in the cross-dispersion direction scales with the instrument
point spread function.  The spectra were flat-fielded and
flux-calibrated by extracting and sky subtracting un-flatfielded
observations of the calibration stars, HR 6348 (IRS-SL), HD 173511
(IRS-LL), and HD 163588 (IRS-HIRES), and dividing these data by their
respective templates (Cohen et al. 2003) to generate a on-dimensional
relative spectral response function (RSRF). The RSRF was then applied
to the TARGET observations to produce the final spectra. The residual
sky was subtracted from the IRS-LORES data using the off-source
observations, which are part of the staring mode AOT. For the
IRS-HIRES data, a spectrum of the zodiacal light on the date of the
observation was generated using the Spitzer Planning and Observation
Tool (SPOT. This uses the zodiacal light model of Reach et al. 2003),
scaled using the IRS-HIRES apertures, and subtracted from the spectra.
The IRS-HIRES spectra were de-fringed in SMART. The spectra were
scaled to match the 25 \um flux densities measured by IRAS.  Sources
lacking a measured IRAS 25 \um flux density were observed with the
IRS' peak-up arrays, allowing us to scale the spectra to match the
source's 22 \um flux density.  The scaling factors, which are
multiplicative, are listed in Table 2.  The ULIRG IRAS 09022-36
required an extremely large scaling factor ($\sim$17) to bring the SH
and 25 \um flux density measured by IRAS into agreement.  We chose not
to scale the data. We are confident in the detection of the S(1) line.
However, the absolute flux calibration of this source is uncertain.

Most of the mid-infrared emission in ULIRGs arises in a region of
order one to a few kiloparsecs in size, which are spatially unresolved
with respect to the IRS slits.\footnotemark[2] \footnotetext[2]{The
  slit sizes for the IRS modules are SL2: 3\farcs6 $\times$ 57$''$,
  SL1: 3\farcs7 $\times$ 57$''$, LL2: 10\farcs5 $\times$ 168$''$, LL1:
  10\farcs7 $\times$ 168$''$, SH: 4\farcs7 $\times$ 11\farcs3, LH:
  11\farcs1 $\times$ 22\farcs3}.  To check this assumption we examined
the IRS-SL2 data. This has the highest spatial resolution with
1.8\arcsec~ per pixel. The IRS-SL2 data was collapsed in the dispersion
direction and a Gaussian was fit to the profile. Our calibration star,
HR 6348 is fit with a Gaussian, with a FWHM of 2.88\arcsec. Only
IRAS~09022-3615, IRAS~12112+0305 and IRAS~14348-1447 show evidence of
significant source extension i.e. FWHM $\ge$ 1.5 $\times$
FWHM$_{star}$. Their FWHM are 3.0, 1.7 and 1.7 times that of the
calibration star, respectively. Both IRAS~12112+0305 and
IRAS~14348-1447 are known doubles with nuclei separated by 2.9 and 3.4
\arcsec, respectively and the 5\arcsec~ Gaussian FWHM corresponds to a
linear diameter of 8 and 9 kpc, respectively. Our observations were
centered on the NE nucleus in IRAS~12112+0305 and the SW nucleus in
IRAS~14348-1447, as these are the dominant nuclei in the near-infrared
and in CO (Evans et al. 2002 and Evans, Surace \& Mazzarella 2000).
For IRAS~09022-3615 there was a pointing problem and the PSF is poorly
sampled. No additional correction to line fluxes has been made to
account for the extended emission.
 
%IRAS09022-3615       8.65534
%IRAS12112+0305       4.95431
%IRAS14348-1447       4.94263
%IRS-SH aperture 4.7\arcsec in the dispersion direction.
%A first order check of this assumption is to
%compare the scale factors applied to IRS-SH and IRS-LH. If a source is
%uniformly extended in both slits the scale factors should differ by
%the slit areas i.e. the IRS-SH scale factor will be approximately five
%times larger than that applied to the IRS-LH data. In Table 2 only
%IRAS 19254-7245 has scale factors appropriate for an extended source.
%note this is prob mispointing issue as the fwhm in SL is IRAS19254-7245 only 40\%
%wider than the cal star

%????these 7 failed in column extraction
%IRAS02115+0226     -71.2800
%IRAS02317-0152     -71.2800
%IRAS02376-0054     -71.2800
%IRAS09346+3911     -71.2800
%IRAS16124+3241     -71.2800
%IRAS17233+3712     -71.2800
%IRAS21293-0154     -71.2800
% note we have 02115 but I checked SL1 extractio and that is
% consistent with a point source so leave as is.

Blending is a problem for the higher rotational lines, which can only
be observed with the IRS-LORES. The S(6) 6.1 \um line, for example, can
be blended with PAH emission at 6.2 $\mu$m, S(5) 6.9 \um line
can be blended with the [Ar~II] 7.0 \um line and PAH emission at 7.0
\ums, and the S(4) 8.0 \um line can be blended with PAH emission at 7.8 \um
and 8.3 \ums. NGC~6240 is the only source in our sample where the
emission is bright enough to be detected in these transitions (Armus
et al. 2006a).

The resulting S(3), S(2), and S(1) \mh line profiles, averaged over
the two slit-nod positions, are shown in the first, second, and third
columns of Figure 1, respectively. The S(0) detections, which
indicates a somewhat cooler \mh component, are shown in Figure 2,
while the S(7) detections, which arise from a hotter \mh component,
are shown in Figure 3. In Table 3 we list the line fluxes and
upper-limits. We find the dominant source of uncertainty in the line
fluxes to be the difference between the fit to the line profiles in
the two slit-nod positions. Upper limits are calculated using the
residuals (rms) from a 0-order fit to the continuum. We take the 3
$\sigma$ upper limit to be 3 $\times$ rms $\times$ FWHM.  No
extinction correction has been applied to the observed line fluxes.
Determining the correct extinction factor is inherently difficult, and
we address this in \S 3.2, where we discuss the location of the warm
molecular hydrogen gas.  We calculate a lower limit to the fraction of
the total far-infrared luminosity, L$_{\rm IR}$\footnotemark[3],
radiated in the \mh for those sources that have at least one measured
emission line flux. The lower limit is the sum of the line
luminosities that were fit with a Gaussian, divided by L$_{\rm IR}$.
\footnotetext[3]{ L$_{\rm IR}$ = L$_{\rm 8 - 1000 ~\mu m}$ is defined
  to be 4$\pi$~d$^{2}_{\rm L}$~F$_{\rm IR}$ in units of L$_{\odot}$,
  where d$_{\rm L}$ is the luminosity distance, and F$_{\rm IR}$ = 1.8
  $\times$ 10$^{-14}$ (13.48~F$_{\rm 12 \mu m}$ + 5.16~F$_{\rm 25 \mu
    m}$ + 2.58~F$_{\rm 60 \mu m}$ + F$_{\rm 100 \mu m}$) in W~m$^{-2}$
  (Sanders and Mirabel 1996).  }Typically $\sim$ 0.01 \% of the total
far-infrared luminosity is radiated in the \mh rotational lines.

\section{Results}

Before deriving the physical properties of the warm molecular hydrogen
we will take a quick look at the sample to check for systematic
offsets and biases. All seventy-seven ULIRGs were observed
using IRS-LORES. However, the brighter sources were also observed in
IRS-HIRES, fifteen with IRS-LH and thirty-seven with both IRS-SH and
IRS-LH.  We find no significant difference in the \mh detection rate
for these IRS-LORES and IRS-HIRES observations. For IRS-HIRES the
detection rate is 35/50 (70$\pm$12\%) as compared to 51/77
(66$\pm$9\%) for IRS-LORES observations.  Twenty-five sources have
both IRS-LORES and IRS-HIRES measurement of the S(1) line. On average
the IRS-HIRES measurement is 90$\pm$31 \% of the IRS-LORES value.
Twenty-three sources have both IRS-LORES and IRS-HIRES measurements of
the S(3) line. Two sources, IRAS~08311-2459 and IRAS~19254-7245, 
have large line flux uncertainties.
The remaining twenty-one sources have an average IRS-HIRES measurement
that is 100$\pm$33 of the average IRS-LORES line flux. There is
no evidence for a significant systematic difference between high- and
low-resolution line flux measurements.

The sample contains a mixture warm and cold sources.  Eighteen of the
twenty-nine warm ULIRGs were detected in \mh (62 $\pm$15\%), whereas
forty-one of the forty-eight cold ULIRGs were detected (85 $\pm$13\%).
There is thus no significant statistical difference in the detection rate
for the two sub-populations.

The likelihood for detecting \mh is flux limited. There are forty
sources in our sample that have an IRAS 60 \um flux density $ \le 2$
Jy. This includes the eighteen sources in the sample for which we do
not detect any molecular hydrogen emission and twelve of the sixteen
sources for which we only detect a single emission line.  This
suggests a detection limit based on our choice of integration time. We
detect molecular hydrogen in all sources with 60 \um flux densities $>
2$ Jy, irrespective of the warm/cold classification.

\subsection{Warm \mh Gas}

To derive the mass of warm molecular hydrogen we assume that all of
our sources are spatially unresolved by the IRS and that the emission
is optically thin. The critical densities of the J = 2, 3, and 4 levels
are relatively low (n$_{\rm cr}< 10^3$ cm$^{-3}$) and we assume that
the populations are in LTE. Adopting an ortho-to-para ratio of 3, we
constructed an excitation diagram for each source.\footnotemark[4]
\footnotetext[4]{The complete set of excitation diagrams is available at
http://isc.astro.cornell.edu/$\sim$sjuh/ulirgs/h2/excitation-diagrams.html.}  
This is simply the natural logarithm of the number of molecules
divided by the statistical weight in the upper level of each
transition versus the energy level. If the \mh is characterized by a
single temperature, the data will lie on a straight line, with the
excitation temperature (T$_{\rm ex}$) being the reciprocal of the slope. 
The excitation diagram for IRAS~00188-0856 is shown in Figure 4. This is a
typical example of single component fit, which is the case for the
majority of our sources. The mass of warm \mh can be derived from the line
luminosity and the excitation temperature. For example, using
luminosity of the S(1) 17.04 \um (ortho) line, the energy of a given
level is given by $E_J = 85 k J(J+1)$, where k is the Boltzmann
constant. The total mass ($M_{Total}$), is $\frac{4}{3} M_o$, where M$_o$
is the mass of gas in the ortho state and $M_o = m_{H_2} N_T$, with
m$_{H_2}$ being the molecular mass of \mh and N$_T$ is the total number
of molecules. However, $N_T$ = $\frac{N_J}{f_J}$, where N$_J$ is the number of
molecules in the J$^{th}$ state, $N_J$ = $\frac{L(J)}{A_{J}\Delta
E_{J}}$.  Here, L(J) is the line luminosity, A$_J$ is the Einstein A
coefficient, $\Delta E_{J} =$ h$\nu_J$, where h is Planck's constant
and $\nu_J$ is the frequency of the emission line. f$_J$ is the
partition function for the J$^{th}$ state, $f_J = \frac{g_J
  e^{-\frac{E_J}{kT_{\rm ex}}}}{\sum_{J',ortho} g_{J'}
  e^{-\frac{E_{J'}}{kT_{\rm ex}}}}$, where g$_J$ is the statistical 
weight for a given state.

We derived an excitation temperature for approximately half of our 
ULIRG sample (43 sources), that had detections in at least two transitions.  
The mean T$_{\rm ex}$ for this sub-sample is 336 $\pm$ 15 K.  Table 3 
also lists an additional sixteen sources in which we only see a single
ortho-molecular hydrogen line.  We used the mean T$_{\rm ex}$ 
to estimate the warm gas mass in these sources. The warm gas masses 
range from 10$^7$ to 10$^9$ \msolars, with an
average value of $\sim$ 2 $\times 10^8$ \msolars.
The warm molecular hydrogen excitation temperatures and masses are given 
in Table 4.

\subsection{Extinction and the Ortho-to-Para Ratio}

It is inherently difficult to derive the correct extinction along a
given line of sight. Using optical depths derived in the mid-infrared
alleviates the problem of extrapolating from a measurement made in the
optical. However, the geometry of the emitting sources and whether the
derived optical depth should be approximated as a foreground screen or
material mixed with the warm gas, either in clumpy or uniform medium,
are all highly uncertain.  Figure 5 shows the combined low-res and
high-res spectrum of Mrk~273 from 6.5 \um to 13.5 \ums. The deep
silicate absorption feature corresponds to an optical depth of $
\tau_{9.7\mu m} \sim$ 2, implying an A$_V \sim 20 - 40$ (Roche \&
Aitken 1985). For molecular hydrogen at a temperature of 380 K in LTE
with an ortho-to-para ratio of 3, the intrinsic S(2)/S(3) line ratio
is 0.56.  If the \mh has the same extinction implied by the depth of
the silicate absorption trough, the line ratio will be in the range
1.05 - 1.95 when the absorbing material in a foreground screen, and
0.74 - 0.90 for a mixture of dust and gas. The observed S(2)/S(3) line
ratio is 0.54 $\pm$ 0.08, which is consistent with the unextincted
value. In Figure 6 we plot T$_{\rm ex}$ versus the observed \mh
S(1)/S(3) line ratio, which is independent of the ortho-to-para ratio.
The data represent the twenty-seven galaxies where at least three J
$\le$ 4 lines are detected in the IRS-HIRES observations. NGC 6240 and
IRAS 12112+0305 are not included in this selection, as their
excitation diagrams are best described by multi-component gas, as we
describe more fully in the next section.  The figure shows that the
majority of the sources have line ratios which are consistent with the
theoretical value for no extinction.  From this we infer that the
absorbing material along the line of sight to the central AGN does not
lie in a screen between us and the warm molecular gas.  If the \mh
emission originates from a massive and dusty circum-nuclear torus,
then the gas and dust are mixed, and the bulk of the emission must
arise from regions with optical depths less than three, as described
by Burton, Hollenback and Tielens (1992).  Alternately, the \mh emission
may also arise in PDRs associated with massive star formation. Higher
angular resolution observations are needed to measure the relative
fractions emitted from a dusty torus and from PDRs.

Using the same twenty-seven sources we can test our assumption that the
ortho-to-para ratio is three. In Figure 7, the dashed line
shows the theoretical \mh S(1)/S(2)
line ratio as a function of T$_{\rm ex}$, and we have over-plotted the 
observed line ratios. The data are consistent with both the assumed
ortho-para-ratio of three with essentially no extinction. In both
Figures 6 and 7, the ULIRG IRAS~15462-0450 shows the largest departure
from the theoretical curves, which we attribute to calibration difficulties.

\subsection{Multi-Temperature \mh Gas}

ULIRGs are complex systems and we would not expect the \mh gas to be
at a single temperature. NGC 6240 (Armus et al. 2006a) is the only
system in our sample where we detect all the S(0) through S(7)
rotational lines.  Its excitation diagram is shown in Figure 8. The
S(3) through S(7) lines are well fit by a hot \mh
component, with T$_{\rm ex}$ = 1327 $\pm$ 36 K. The S(7) line
flux implies a hot molecular hydrogen mass of 
(4.1 $\pm$ 0.8) $\times$ 10$^6$ \msolars. Subtracting this
component's contribution from the S(0) through S(3) lines 
gives a second, cooler, \mh component, with T$_{\rm ex}$ =
292 $\pm$ 6 K. The measured S(0) line flux weights the fit to 
lower temperatures and correspondingly larger \mh masses. The
warm gas mass inferred for the S(0) line flux is (6 $\pm$ 3)
$\times$ 10$^8$ \msolars.

Besides NGC~6240, we detect the S(0) line in only two other sources,
IRAS~12112+0305 and IRAS~13342+3932. Likewise, the S(7) line is detected 
in only six sources, NGC~6240, Arp~220, IRAS~06301-7954, IRAS~12032+1707, 
IRAS~12112+0305, and IRAS~17068+4027 with S/N $>$ 3. Figure 9 shows the
excitation diagrams for these six sources.  When the S(7) line is detected, 
a fit is first made to the S(7) and S(3) lines to determine the
temperature of the hot \mh component. In Arp 220, for example, the 
derived excitation temperature is
1435 $\pm$ 120 K. This component's contribution is then subtracted 
from the S(3) through S(0) lines, so the T$_{\rm ex}$ and mass 
of the cooler \mh component can be estimated. The derived T$_{\rm ex}$ 
and \mh masses are listed in Table 4.
In IRAS~13342+3932 the fit is heavily weighted towards a low
temperature, as only the S(0) and S(1) lines are detected. The
resulting gas mass is an order of magnitude larger than the masses 
calculated for
NGC~6240 and IRAS~12112+0305. A similar decrease of $\sim$ 50-100 K in
the temperature and increase in the mass by a factor of ten is
obtained for both NGC~6240 and IRAS~12112+0305, if only the S(0) and
S(1) lines are used in the second temperature fit. This suggests that our
result for both the mass and temperature in IRAS~13342+3932 is
strongly biased by our non-detection of the S(2) and S(3) lines. Using
the average temperature for the sample (T$_{\rm ex}$ = 
336 $\pm$ 15 K), we infer a mass of $\sim$2$\times$10$^8$ M$_{\odot}$, 
which is probably a better estimate of the
warm \mh gas mass, and more in keeping with the rest of the sample. 
The average hot gas temperature is 1400 $\pm$ 60 K, with an associated 
molecular mass of (3 $\pm$ 2)$\times$10$^6$ \msolars. 
The warm, T$_{\rm ex}$ $\sim$ 340 K \mh component, with
M$_{\rm H_2}$ $\sim$ 10$^8$ - 10$^{9}$ \msolars,
thus represents the bulk of the measured gas mass.

\section{Discussion}

\subsection{Warm and Cold ULIRGs}

Figure 10 shows the derived \mh mass as a function of the IRAS
25 $\mu$m and 60 $\mu$m flux density ratio (F$_{25 \mu m}$/F$_{60 \mu m}$).
The vertical dashed line corresponds to F$_{25 \mu m}$/F$_{60 \mu m}$ = 0.2,
which is used to separate ``cold'' and ``warm'' ULIRGs.
There are no obvious trends. That is, there appears to be no systematic
tendency for starburst or AGN powered ULIRGs to have systematically
larger (or smaller) masses of warm \mh. 

Likewise, there is no clear dependence on far-infrared luminosity.
Figure 11 shows the derived warm \mh mass as a function of the
60 $\mu$m specific luminosity (L$_{\rm 60 \mu m}$), which we define
as 4$\pi$~d$^{2}_{\rm L}$~F$_{\rm 60 \mu m}$ in units of W~Hz$^{-1}$.
Here, d$_{\rm L}$ is the source's luminosity distance.
The twenty-seven ULIRGs in the plot all have z $<$ 0.1, which ensures
that the IRAS 60 $\mu$m flux densities correspond to nearly
identical rest-frame wavelengths. A possible tendency for warm \mh
mass to increase with far-infrared luminosities in excess of 
L$_{\rm 60 \mu m}$ = 3 $\times$ 10$^{25}$ W~Hz$^{-1}$ is significantly
weakened by a number of outliers with higher \mh masses.

The observed S(1) line intensity is typically
$\sim$10$^{-4}$ erg cm$^{-2}$ s$^{-1}$, which is consistent with the 
Burton, Hollenbach, and Tielens (1992) PDR models with density 
$\sim$10$^{4}$-10$^{5}$ cm$^{-3}$ and far-UV fields of 
$\sim$10$^3$-10$^4$ G$_0$. Here, G$_0$ is the local interstellar 
far-UV radiation field, determined to be 1.6 $\times$ 10$^{-3}$ ergs
s$^{-1}$ cm$^{-2}$ (Habing 1968). This is consistent with PDRs being a
large contributor to the warm \mh emission. Of course there can also be
a significant shock component in some ULIRGs. Constraining the
contribution of shock excitation will require observations of
ro-vibrational transitions in the near-infrared.

\subsection{Fractional Warm Gas Mass}

In Table 4 we have listed the cold molecular hydrogen mass derived
from $^{12}$CO(1-0) observations of thirty ULIRGs from the literature.
These masses range from 10$^8$ to 10$^{11}$ M$_{\odot}$, with an
average cold M$_{\rm H_2}$ $\sim$ 4 $\times$ 10$^{10}$ M$_{\odot}$.
This is calculated using the standard Milky Way M$_{\rm H_2}$/L$_{\rm
  CO}$ ratio of 4.6 M$_{\odot}$/K km s$^{-1}$ pc$^{2}$ determined from
observations of Galactic molecular clouds (Solomon et al. 1987).
However, as stressed by Maloney \& Black (1988), because the CO(1-0)
line intensity scales as a modest power of the gas
temperature\footnotemark[5] \footnotetext[5]{I$_{\rm CO}$ $\propto$
  T$_{\rm kin}^{1.3}$ for T$_{\rm kin}$ $\la$ 30 K, and I$_{\rm CO}$
  $\propto$ T$_{\rm kin}$ for T$_{\rm kin}$ $>$ 30 K.}, the molecular
hydrogen masses may be {\em overestimated} by factors of 3-5 in
regions of intense star formation or AGN dominance. The density and
filling factor of the $^{12}$CO(1-0) emitting regions also affect the
measured CO intensity in ways that are difficult to take into account
a priori. Solomon et al.  (1997) argue that the standard conversion
factor is inappropriate for ULIRGs, as the derived masses often exceed
the dynamical masses.  They argue that in the extreme environments
found in the central regions of ULIRGs, the CO emission originates not
from self-gravitating clouds, but from an inter-cloud medium, which is
bound by the potential of the galaxy or molecular gas pressure,
resulting in a smaller M$_{\rm H_2}$/L$_{\rm CO}$ of $\sim$ 1.4
M$_{\odot}$/K km s$^{-1}$ pc$^{2}$.  All of this suggests that the
cold molecular masses in Table 4 should be viewed with considerable
caution, and likely represent overestimates of the true M$_{H_2}$.
Indeed, use of the $\alpha$ advocated by Solomon et al. (1997) for
their ULIRG sample, would both lower the cold H$_{2}$ mass and
increase the warm H$_{2}$ gas fraction by a factor of $\sim$3. However
the warm H$_{2}$ gas remains a small fraction of the total H$_{2}$
mass.

We have \mh detections for twenty-eight of these sources. Some of the
sources do have a warm gas fraction that scales with warm gas mass,
but the scatter in the data is large and there is no strong trend.
The median fractional warm gas mass is 0.3\% of the cold gas mass. Figure
12 shows the warm gas fraction as a function of 
F$_{25 \mu m}$/F$_{60 \mu m}$. Rigopoulou et al. (2002) derived an 
average gas temperature of $\sim$150 K for their sample of local 
starbursts and AGN, and found the warm
gas fraction to as large as 10\% in starbursts and up to 35\% (with
a large scatter) in Seyferts. In nearly half of their sample they did
not detect the S(0) line, and the derived temperature is $\sim$300 K.
The difference between the average warm gas fractions in our ULIRG
sample and the local starburst/AGN sample is heavily weighted by 
the adopted gas temperature, and whether or not the S(0) line is 
detected. It is likely that the warm gas fraction is similar in both. 
More sensitive observations of the S(0) line in ULIRGs are needed 
to fully constrain the $\sim$150 K component.

Figure 12 also shows that only 8/28 of ULIRGs with both cold and
warm \mh masses have F$_{25 \mu m}$/F$_{60 \mu m}$ $\ge$ 0.2, and
only one has F$_{25 \mu m}$/F$_{60 \mu m}$ $\ge$ 0.4. As a result, 
we are unable to say conclusively if there is a tendency for the 
warm \mh mass to depend on dust temperature, i.e., on AGN dominance. 
While the position of Mrk~463 in the upper-right of the figure is suggestive
of such a trend, it may simply be an outlier, like the ``cold''
IRAS~12112+0305 near the top-left. No increase in the warm gas fraction 
is measured as a function of either increasing IRAS 25 $\mu$m or 60 $\mu$m 
specific luminosity.

\section{Conclusions}

We have used Spitzer's IRS to obtain spectra from
5.0$-$38.5 $\mu$m for a large sample of ULIRGs. This
sample is bright (0.1 $<$ F$_{60 \mu m} \le$ 104 Jy) and have redshifts
between 0.02 and 0.93. We have detected line emission
from pure rotational transitions of molecular hydrogen in
77$\%$ of the sample, and in all ULIRGs with F$_{\rm 60 \mu m}$
$>$ 2 Jy. The majority of these sources are modeled with a single 
temperature \mh component, with an average excitation temperature
(T$_{\rm ex}$) of 336 $\pm$ 15 K. The ULIRGs contain large 
quantities of warm molecular hydrogen. We find an average mass
 2 $\times$ 10$^8$ M$_{\odot}$. Large optical depths, 
as inferred from the depth of the silicate absorption trough at 9.7
$\mu$m, are not observed along the line of
site to the warm \mh gas. If the gas originates in a central dusty
torus, the medium must be clumpy. The warm \mh mass does not appear
to depend on the IRAS 25 $\mu$m to 60 $\mu$m flux density ratio,
which is used to classify ULIRGs into ``warm'' (AGN dominated) and
``cold'' (starburst dominated) categories. Similarly, the warm
\mh mass does not scale with the specific luminosity at 25 $\mu$m
or 60 $\mu$m. A possible common origin for the gas is from 
photo-dissociation regions associated with massive star formation.

The single temperature gas model, fitted to the S(1) through S(3) lines, 
is a simplification and the gas is likely to have multiple temperature
components.  Detection of the S(0) line will typically lower the derived
temperature in the single component fit by 50-100 K, leading to a
larger estimated gas mass. This higher gas mass cannot be ruled out
in the bulk of our sample for which we do not detect the S(0) line. 
A hot \mh gas component is inferred when the S(7) line is detected. 
This line is detected in six ULIRGs, and is modeled as a
T$_{\rm ex}$ = 1400 $\pm$ 60 K \mh component, with a relatively 
small mass (M$_{\rm H_2}$ $\sim$ 3 $\times$ 10$^6$ M$_{\odot}$).

Using cold molecular gas estimates from $^{12}$CO observations in the
literature, we find that the warm \mh mass is typically less than one
percent of the cold \mh mass.  The average warm gas fraction could be
a few tens of percent, but this requires verification through the
detection of the S(0) 28.22 $\mu$m line, which is difficult using the
IRS as the line is situated in a low-sensitivity region of the
detector array, and detection of the $^{12}$CO line. In general,
ULIRGs have a warm mass fraction that is comparable to that found in
local starburst and Seyfert galaxies (Rigopoulou et al. 2002).

\acknowledgments This work is based [in part] on observations made
with the Spitzer Space Telescope, which is operated by the Jet
Propulsion Laboratory, California Institute of Technology under NASA
contract 1407. Support for this work was provided by NASA through
Contract Number 1257184 issued by JPL/Caltech.

We with to thank Paul Goldsmith, David Hollenbach, Alberto
Noriego-Crespo and Terry Herter for useful discussions concerning the
excitation of molecular hydrogen, Bill Reach for the Zodiacal light
model, and James Houck for making the IRS possible.

\references 

\reference{} Adams , T. F. \& Weedman, D. W. 1972, ApJL, 173, 109

\reference{} Armus et al. 2004, ApJS, 154, 178.

\reference{} Armus et al. 2006a, ApJ, in press.

\reference{} Armus et al. 2006b, ApJ in press.

\reference{} Arakelian, M. A., Dibai, E. A., Esipov, V. F.
\& Markarian, B. E. 1971, Astrofizika, 7, 177.

\reference{} Burton, M. G., Hollenbach, D. J., \& Tielens, A. G. 1992,
\apj, 399, 563

\reference{} Cohen, M., Megeath, T.G., Hammersley, P.L., Martin-Luis, F., 
\& Stauffer, J. 2003, \aj, 125, 2645

\reference{} Cutri, R. et al. 2003, 2MASS All-Sky Catalog of Point Sources, VizieR On-line
Data Catalog: II/246. Originally published in: University of
Massachusetts and Infrared Processing and Analysis Center,
(IPAC/California Institute of Technology) (2003)

\reference{} Elbaz, D., \& Cesarsky, C.~J. 2003, Science, 300, 270

\reference{} Evans, A. S., Surace, J.A. \& Mazzarella, J. M. 2000, 
ApJL, 529, 85

\reference{} Evans, A. S., Mazzarella, J. M., Surace, J.A. \& Sanders,
D. B. 2002, ApJ, 580, 749

\reference{} Habart, E. Walmsley, C. M., Verstraete, L., Cazaux, S.,
Maiolino, R.,Cox, P., Boulanger, F.,Pineau des Forets, G. 2005, SSR,
119,71

\reference{} Habing, H. J. 1968, Bull. Astr. Inst. Netherlands, 19, 421.

\reference{} Higdon, S. J. U. et al. 2004, PASP, 116, 975 

\reference{} Houck, J. R.  et al., 2004 ApJS, 154, 18

\reference{} Kim, D. -C. \& Sanders, D. B. 1998, ApJS, 119, 41

\reference{} Kormendy, J. \& Sanders, D. B. 1992, ApJL, 390, 53

\reference{} Lonsdale, C. J., Lonsdale, C. J., Smith, H. E., 
Diamond, P. J. 2003, ApJ, 592, 804	

\reference{} Lutz, D., Sturm, E.;, Genzel, R., Spoon, H. W. W., 
Moorwood, A. F. M., Netzer, H. \& Sternberg, A. 2003, A\&A, 409, 867

\reference{} Maloney, P. \& Black, J. H. 1988, \apj, 325, 389

%co paper 6
\reference{} Mirabel, I. F., Booth, R. S. Garay, G., Johansson, L. E.
B. \& Sanders, D. B. 1990, A\&A, 236, 327

\reference{} Moorwood, A. F. M. 1996, SSRv, 77, 303

\reference{} Neugebauer, G. et al. 1984, ApJL, 278 1

\reference{} Reach, W.T., Morris, P., Boulanger, F., \& Okumura, K.
2003, Icarus, 164, 384.

\reference{} Roche, P. F. \& Aitken, D. K. 1985, MNRAS, 215, 425

%co paper 1
\reference{} Rigopoulou, D., Lawrence,A., White, G. J.,Rowan-Robinson,
M. \& Church, S. E. 1996, A\&A, 305, 747

\reference{} Rigopoulou, D., Kunze, D., Lutz, D., Genzel, R. 
\& Moorwood, A. F. M 2002, A\&A, 389, 374

\reference{} Sanders, D. B. \& Mirabel, I. F. 1996, ARA\&A, 34, 749

\reference{}  Sanders, D. B., Scoville, N. Z. \& Soifer, B. T. 1991, ApJ 370, 158 

\reference{}  Sanders, D. B., Soifer, B. T., Elias, J. H., Madore, B. F., 
Matthews, K., Neugebauer, G., \& Scoville, N. Z. 1988, ApJ, 325, 74

\reference{} Soifer, B. T., Sanders, D. B., Madore, B. F., Neugebauer, G., 
Danielson, G. E., Elias, J. H., Lonsdale, C. J. 
\& Rice, W. L. 1987, ApJ, 320, 238

%co paper 2
\reference{} Solomon, P. M., Downes, D., Radford, S. J. E. \& Barrett,
J. W. 1997, ApJ, 478, 144

\reference{} Solomon, P. M., Rivolo, A. R., Barrett, J. W. \& Yahil,
A. 1987, ApJ, 319, 730

\reference{} Stanford, S. A., Stern, D., van Breugel, W. \& De Breuck,
C. 2000, ApJS, 131,185

\reference{} Strauss, M. A., Huchra, J. P., Davis, M., Yahil, A., 
Fisher, K B. \& Tonry, J. 1992, ApJS, 83, 29

\reference{} Sturm et al. 1996, A\&A, 315L, 133

\reference{} Valentijn, E. A., van der Werf, P. P., de Graauw, T. 
\& de Jong, T 1996, A\&A, 315L, 145

\reference{} Van der Werf, P. P., Genzel, R.  Krabbe, A. et al. 1993,
ApJ 405 522

\reference{} Werner, M. W., et al. 2004, ApJS, 154, 1

\clearpage
%
% table 1
%
%
\begin{deluxetable}{lcccc}
\tabletypesize{\scriptsize}
\tablecolumns{5} 
\tablewidth{0pc}
\tablecaption{ULIRG Sample}
\tablehead{
\colhead{Object}&
\colhead{z}&
\colhead{D$_{L}$\tablenotemark{a}  }&
\colhead{L$_{IR}$\tablenotemark{b}}&
\colhead{  $\frac{ F_{25 \mu m}}{F_{60 \mu m}}$  }\\
\colhead{}&
\colhead{}&
\colhead{( Mpc )}&
\colhead{ (10$^{12}$L$_{\odot}$)}&
\colhead{}
}
\startdata
           3C273  &0.1584  & 749.3  & 5.74  &0.52 \\
         Arp~220  &0.0181  &  77.6  & 1.45  &0.08 \\
 IRAS~00188-0856  &0.1286  & 596.3  & 2.56  &0.14 \\
 IRAS~00275-0044  &0.2420  &1203.8  & 2.62  &0.37 \\
 IRAs~00275-2859  &0.2792  &1418.1  & 3.16  &0.25 \\
 IRAS~00397-1312  &0.2617  &1316.5  & 6.46  &0.22 \\
 IRAS~00406-3127  &0.3424  &1797.9  & 6.25  &0.12 \\
 IRAS~00476-0054  &0.7270  &4462.9  & 8.32  &0.11 \\
 IRAS~01003-2238  &0.1177  & 541.9  & 1.95  &0.29 \\
 IRAS~01199-2307  &0.1560  & 736.6  & 2.32  &0.10 \\
 IRAS~02115+0226  &0.4000  &2160.1  & 9.65  &0.49 \\
 IRAS~02433+0110  &0.7980  &5009.1  & 8.91  &0.23 \\
 IRAS~03521+0028  &0.1522  & 716.7  & 3.65  &0.09 \\
 IRAS~04114-5117  &0.1246  & 576.2  & 1.85  &0.04 \\
 IRAS~04313-1649  &0.2680  &1352.9  & 4.48  &0.07 \\
 IRAS~05189-2524  &0.0426  & 185.7  & 1.43  &0.25 \\
 IRAS~06035-7102  &0.0795  & 356.3  & 1.66  &0.11 \\
 IRAS~06206-6315  &0.0924  & 418.2  & 1.69  &0.07 \\
 IRAS~06301-7934  &0.1564  & 738.7  & 2.45  &0.05 \\
 IRAS~06361-6217  &0.1596  & 755.3  & 1.49  &0.10 \\
 IRAS~07598+6508  &0.1488  & 699.5  & 3.37  &0.31 \\
 IRAS~08311-2459  &0.1004  & 457.0  & 3.31  &0.22 \\
 IRAS~08572+3915  &0.0584  & 257.6  & 1.37  &0.23 \\
 IRAS~09022-3615  &0.0596  & 263.6  & 1.91  &0.10 \\
 IRAS~09463+8141  &0.1547  & 729.8  & 2.03  &0.05 \\
 IRAS~10091+4704  &0.2460  &1226.5  & 4.55  &0.07 \\
 IRAS~10398+3247  &0.6330  &3762.8  & 6.03  &0.04 \\
 IRAS~10565+2448  &0.0431  & 188.2  & 1.09  &0.09 \\
 IRAS~11119+3257  &0.1890  & 911.1  & 4.51  &0.22 \\
 IRAS~12018+1941  &0.1686  & 802.7  & 3.18  &0.21 \\
 IRAS~12032+1707  &0.2170  &1063.8  & 3.92  &0.17 \\
 IRAS~12071-0444  &0.1284  & 595.2  & 2.30  &0.22 \\
 IRAS~12112+0305  &0.0727  & 324.2  & 2.12  &0.06 \\
 IRAS~12514+1027  &0.3000  &1541.0  & 5.53  &0.27 \\
 IRAS~13120-5453  &0.0308  & 133.0  & 1.73  &0.07 \\
 IRAS~13218+0552  &0.2051  & 998.4  & 5.14  &0.34 \\
 IRAS~13342+3932  &0.1793  & 859.2  & 1.96  &0.02 \\
 IRAS~13352+6402  &0.2366  &1173.3  & 3.64  &0.08 \\
 IRAS~13451+1232  &0.1220  & 563.3  & 1.97  &0.35 \\
 IRAS~14348-1447  &0.0825  & 370.6  & 2.24  &0.07 \\
 IRAS~14378-3651  &0.0676  & 300.7  & 1.35  &0.08 \\
 IRAS~14537+1950  &0.6400  &3814.0  &10.49  &0.57 \\
 IRAS~14548+3349  &0.4430  &2440.0  & 2.82  &0.18 \\
 IRAS~15001+1433  &0.1627  & 771.7  & 2.79  &0.09 \\
 IRAS~15206+3342  &0.1244  & 575.3  & 1.73  &0.20 \\
 IRAS~15250+3609  &0.0554  & 244.0  & 1.13  &0.18 \\
 IRAS~15462-0450  &0.0998  & 453.8  & 1.59  &0.15 \\
 IRAS~16124+3241  &0.7100  &4334.3  & 4.57  &0.31 \\
 IRAS~16334+4630  &0.1910  & 921.9  & 2.89  &0.08 \\
 IRAS~17068+4027  &0.1791  & 858.2  & 2.43  &0.09 \\
 IRAS~17179+5444  &0.1475  & 692.5  & 1.98  &0.15 \\
 IRAS~17208-0014  &0.0430  & 187.8  & 2.50  &0.05 \\
 IRAS~17233+3712  &0.6890  &4176.6  &10.51  &0.25 \\
 IRAS~17463+5806  &0.3090  &1594.8  & 2.69  &0.06 \\
 IRAS~18030+0705  &0.1458  & 683.8  & 1.90  &0.30 \\
 IRAS~18443+7433  &0.1347  & 627.4  & 2.08  &0.09 \\
 IRAS~19254-7245  &0.0617  & 273.1  & 1.21  &0.23 \\
 IRAS~19297-0406  &0.0857  & 386.1  & 2.56  &0.08 \\
 IRAS~19458+0944  &0.0999  & 454.6  & 2.32  &0.07 \\
 IRAS~20037-1547  &0.1919  & 926.7  & 3.66  &0.17 \\
 IRAS~20087-0308  &0.1057  & 482.5  & 2.80  &0.05 \\
 IRAS~20100-4156  &0.1296  & 601.3  & 4.48  &0.07 \\
 IRAS~20414-1651  &0.0871  & 392.5  & 1.70  &0.08 \\
 IRAS~20551-4250  &0.0427  & 186.2  & 1.10  &0.15 \\
 IRAS~21272+2514  &0.1508  & 709.6  & 1.40  &0.36 \\
 IRAS~22491-1808  &0.0773  & 346.0  & 1.54  &0.10 \\
 IRAS~23128-5919  &0.0446  & 194.9  & 1.07  &0.15 \\
 IRAS~23129+2548  &0.1791  & 858.2  & 3.13  &0.07 \\
 IRAS~23365+3604  &0.0645  & 286.0  & 1.47  &0.11 \\
 IRAS~23498+2423  &0.2120  &1036.3  & 2.99  &0.20 \\
 IRAS~23529-2119  &0.4303  &2356.5  & 4.68  &0.48 \\
         Mrk~231  &0.0422  & 184.2  & 3.70  &0.25 \\
         Mrk~273  &0.0378  & 164.3  & 1.42  &0.10 \\
        Mrk~463E  &0.0504  & 221.0  & 0.60  &0.72 \\
       Mrk~1014   &0.1631  & 773.6  & 4.12  &0.24 \\
        NGC~6240  &0.0245  & 105.4  & 0.69  &0.15 \\
        UGC~5101  &0.0400  & 174.3  & 1.00  &0.09 \\
\enddata

\tablenotetext{a}{The luminosity distance in Mpc.}
\tablenotetext{b}{The 8-1000 $\mu$m far-infrared luminosity, defined as
L$_{\rm IR}$ = 4$\pi$ d$^{2}_{L}$~F$_{\rm IR}$ in units of L$_{\odot}$, with
F$_{\rm IR}$ =1.8 $\times$ 10$^{-14}$ (13.48 F$_{\rm 12 \mu m}$ + 
5.16 F$_{\rm 25 \mu m}$ + 2.58 F$_{\rm 60 \mu m}$ + F$_{\rm 100 \mu m}$), 
in units of W~m$^{-2}$ (Sanders and Mirabel 1996). }

\end{deluxetable}
\clearpage

%%%%%%%%%%%%%%%%%%%%Table 2

\begin{deluxetable}{ccccccccccc}
\tabletypesize{\scriptsize}
\rotate
\tablecolumns{11}
\tablewidth{0pc}
\tablecaption{Observational Parameters}
\tabletypesize{\footnotesize}
\tablehead{
\colhead{Object}&
\colhead{ RA}& 
\colhead{Dec}& 
\colhead{Date Obs.}& 
\colhead{Exec. Time} &
\multicolumn{6}{c}{Scale Factor to IRAS 25 \um flux density} \\
\colhead{}&
\colhead{(J2000)}&
\colhead{(J2000)}&
\colhead{}& 
\colhead{(min)}& 
\colhead{SL2} & 
\colhead{SL1} & 
\colhead{LL2} & 
\colhead{LL1} & 
\colhead{SH} & 
\colhead{ LH} }
\startdata 
3C273&12:29: 6.67&  2: 3: 8.10&2005-06-07& 21.7&1.01&1.03&0.97&0.98&0.96&0.78
\\
Arp220&15:34:57.24& 23:30:11.70&2004-02-29& 42.8&1.00&1.14&0.97&0.92&1.08&1.02
\\
%3C273 BAD??&12:29: 6.67&  2: 3: 8.10&2004-01-06& 42.2&1.75&2.03&1.46&1.43&
%\nodata&1.14\\
IRAS00188-08& 0:21:26.48& -8:39:27.10&2003-12-17& 57.0&0.87&0.93&0.84&0.85&
0.60&0.56\\
IRAS00275-00& 0:30: 9.09& -0:27:44.40&2005-07-08& 47.2&2.33&2.16&1.88&1.70&
\nodata&\nodata\\
IRAS00275-28& 0:30: 4.20&-28:42:25.40&2003-12-17& 28.6&0.87&0.92&0.80&0.81&
\nodata&\nodata\\
IRAS00397-13& 0:42:15.50&-12:56: 3.50&2004-01-04& 57.0&0.97&0.97&0.90&0.92&
0.70&0.54\\
IRAS00406-31& 0:43: 3.14&-31:10:49.70&2005-07-11& 47.2&0.95&0.99&0.93&0.94&
\nodata&\nodata\\
IRAS00476-00& 0:50: 9.81& -0:39: 1.00&2004-07-18& 67.8&0.82&0.84&0.75&0.69&
\nodata&0.19\\
IRAS01003-22& 1: 2:49.94&-22:21:57.30&2004-01-04& 44.3&0.96&1.04&0.85&0.82&
0.87&0.70\\
IRAS01199-23& 1:22:20.84&-22:51:57.30&2004-07-18& 30.4&1.08&1.16&0.96&0.92&
\nodata&\nodata\\
%IRAS01368+01& 1:39:27.18&  1:15:47.20&2004-07-16& 67.8&1.03&0.07&1.11&1.10&
%\nodata&0.16\\
IRAS02115+02& 2:14:10.32&  2:39:59.80&2005-02-07& 23.7&0.18&0.74&0.66&0.75&
\nodata&\nodata\\
%IRAS02317-01& 2:34:21.90& -1:39: 0.70&2005-02-11& 48.0&\nodata&\nodata&\nodata
%&\nodata&\nodata&\nodata\\
%IRAS02376-00& 2:40: 8.58& -0:42: 3.60&2005-02-12& 48.5&0.53&0.03&0.76&0.93&
%\nodata&\nodata\\
IRAS02433+01& 2:45:55.36&  1:23:28.40&2005-01-12& 43.0&1.00&1.00&0.99&0.98&
\nodata&\nodata\\
%IRAS03158+42& 3:19:12.60& 42:38:28.00&2004-03-01& 43.0&\nodata&\nodata&\nodata
%&\nodata&\nodata&\nodata\\
IRAS03521+00& 3:54:42.15&  0:37: 2.00&2004-02-27& 57.2&0.90&0.90&0.80&0.81&
0.42&0.48\\
IRAS04114-51& 4:12:44.92&-51: 9:34.20&2004-08-11& 28.5&0.98&1.02&0.97&1.02&
\nodata&\nodata\\
IRAS04313-16& 4:33:37.08&-16:43:31.50&2004-03-01& 35.8&0.94&1.14&0.91&0.92&
\nodata&\nodata\\
IRAS05189-25& 5:21: 1.41&-25:21:45.50&2004-03-22& 41.2&1.05&1.09&1.00&1.01&
1.07&1.07\\
IRAS06035-71& 6: 2:53.63&-71: 3:11.90&2004-04-14& 42.9&1.37&1.47&1.17&1.07&
1.73&1.07\\
IRAS06206-63& 6:21: 0.80&-63:17:23.20&2004-04-16& 44.1&0.91&0.96&0.84&0.78&
1.05&0.79\\
IRAS06301-79& 6:26:42.20&-79:36:30.40&2004-08-11& 28.5&1.00&1.09&0.92&0.95&
\nodata&\nodata\\
IRAS06361-62& 6:36:35.71&-62:20:31.80&2004-08-11& 30.4&0.94&0.97&0.96&0.98&
\nodata&\nodata\\
IRAS07598+65& 8: 4:30.46& 64:59:52.90&2004-02-29& 43.2&1.08&1.11&1.01&0.98&
1.05&0.92\\
IRAS08311-24& 8:33:20.47&-25: 9:33.10&2004-04-14& 42.9&\nodata&\nodata&\nodata
&\nodata&1.68&1.16\\
IRAS08572+39& 9: 0:25.38& 39: 3:54.30&2004-04-15& 42.3&1.05&1.13&1.00&0.91&
1.05&0.89\\
IRAS09022-36& 9: 4:12.69&-36:27: 1.70&2005-06-07& 37.8&\nodata&\nodata&\nodata
&\nodata&\nodata&\nodata\\
%IRAS09346+39& 9:37:41.66& 38:57:52.60&2004-11-16& 47.4&\nodata&\nodata&\nodata
%&\nodata&\nodata&\nodata\\
IRAS09463+81& 9:53: 0.09& 81:27:28.20&2004-03-23& 28.5&1.27&1.29&0.94&0.95&
\nodata&\nodata\\
IRAS10091+47&10:12:16.74& 46:49:42.90&2004-04-19& 28.5&0.87&0.88&0.84&0.85&
\nodata&\nodata\\
IRAS10398+32&10:42:40.81& 32:31:31.00&2004-12-08& 42.9&0.85&0.81&0.02&0.76&
\nodata&\nodata\\
IRAS10565+24&10:59:18.14& 24:32:34.30&2004-05-12& 42.8&1.22&1.32&1.11&1.03&
1.12&0.97\\
IRAS11119+32&11:14:38.88& 32:41:33.10&2004-05-11& 52.0&1.20&1.57&1.00&1.01&
1.10&0.86\\
%3C273 BAD??&12:29: 6.67&  2: 3: 8.10&2004-01-06& 42.2&\nodata&\nodata&\nodata&
%\nodata&\nodata&\nodata\\
IRAS12018+19&12: 4:24.53& 19:25: 9.80&2004-05-15& 51.9&1.06&1.10&0.92&0.92&
0.87&0.76\\
IRAS12032+17&12: 5:47.73& 16:51: 8.20&2004-01-04& 28.6&1.22&1.27&1.00&0.95&
\nodata&\nodata\\
IRAS12072-04&12: 9:45.12& -5: 1:13.90&2004-01-06& 44.3&1.16&1.23&1.11&1.12&
0.99&0.88\\
IRAS12112+03&12:13:46.05&  2:48:41.30&2004-01-04& 44.0&1.33&1.37&1.24&1.25&
1.12&0.93\\
IRAS12514+10&12:54: 0.82& 10:11:12.40&2005-02-06& 62.7&0.96&0.98&0.92&0.91&
0.77&0.60\\
IRAS13120-54&13:15: 6.40&-55: 9:23.30&2004-03-02& 40.1&1.38&1.49&1.07&1.02&
1.24&1.09\\
IRAS13218+05&13:24:19.81&  5:37: 4.60&2004-07-17& 51.7&0.97&0.99&0.91&0.91&
\nodata&0.60\\
IRAS13342+39&13:36:24.07& 39:17:30.10&2004-05-12& 64.1&1.07&1.07&0.97&0.94&
\nodata&0.88\\
IRAS13352+64&13:36:51.15& 63:47: 4.70&2005-03-20& 47.2&0.99&0.96&0.94&0.94&
\nodata&\nodata\\
IRAS13451+12&13:47:33.36& 12:17:24.20&2004-01-07& 42.9&1.21&1.29&1.19&1.21&
1.11&1.02\\
IRAS14348-14&14:37:38.27&-15: 0:24.60&2004-02-07& 44.4&1.07&1.25&1.17&1.13&
0.99&0.81\\
IRAS14378-36&14:40:58.90&-37: 4:33.00&2004-03-02& 43.2&1.70&1.51&1.25&1.19&
1.62&1.02\\
IRAS14537+19&14:56: 4.43& 19:38:45.70&2004-07-14& 67.5&0.85&0.85&0.77&0.74&
\nodata&0.22\\
IRAS14548+33&14:56:58.43& 33:37:10.00&2005-03-17& 43.1&0.63&0.96&0.91&0.84&
\nodata&\nodata\\
IRAS15001+14&15: 2:31.94& 14:21:35.30&2004-06-27& 56.9&1.09&1.25&0.96&0.95&
\nodata&0.79\\
IRAS15206+33&15:22:38.12& 33:31:36.10&2004-06-24& 51.8&1.01&1.07&0.83&0.83&
\nodata&0.78\\
IRAS15250+36&15:26:59.40& 35:58:37.50&2004-03-04& 43.4&1.03&1.00&0.97&0.92&
1.01&0.92\\
IRAS15462-04&15:48:56.80& -4:59:33.70&2004-03-02& 52.1&1.23&1.35&1.04&1.00&
0.96&0.83\\
IRAS16124+32&16:14:22.11& 32:34: 3.70&2005-03-19& 48.0&0.41&0.92&0.88&0.87&
\nodata&\nodata\\
IRAS16334+46&16:34:52.37& 46:24:53.00&2004-03-04& 28.8&0.83&0.35&0.65&0.63&
\nodata&\nodata\\
IRAS17068+40&17: 8:32.12& 40:23:28.20&2004-04-16& 30.4&0.89&0.90&0.90&0.93&
\nodata&\nodata\\
IRAS17179+54&17:18:54.23& 54:41:47.30&2004-04-17& 66.6&1.14&1.25&1.08&1.04&
1.18&1.07\\
IRAS17208-00&17:23:21.93& -0:17: 0.40&2004-03-27& 42.2&1.33&1.35&1.17&1.07&
1.19&1.03\\
IRAS17233+37&17:25: 7.40& 37: 9:32.10&2004-06-06& 67.6&0.25&1.12&1.12&0.95&
\nodata&0.94\\
IRAS17463+58&17:47: 4.76& 58: 5:22.10&2004-04-17& 32.8&0.96&0.96&0.91&0.92&
\nodata&\nodata\\
IRAS18030+07&18: 5:32.50&  7: 6: 9.00&2004-03-25& 28.5&0.87&0.93&0.78&0.75&
\nodata&\nodata\\
IRAS18443+74&18:42:54.80& 74:36:21.00&2004-03-05& 30.7&1.01&0.97&0.94&0.95&
\nodata&\nodata\\
IRAS19254-72&19:31:21.55&-72:39:22.00&2005-05-30& 37.7&\nodata&\nodata&\nodata
&\nodata&6.62&1.31\\
IRAS19297-04&19:32:21.25& -3:59:56.30&2004-05-13& 42.7&0.95&0.97&0.84&0.82&
\nodata&0.71\\
IRAS19458+09&19:48:15.70&  9:52: 5.00&2004-04-18& 30.4&1.15&1.14&1.05&0.96&
\nodata&\nodata\\
IRAS20037-15&20: 6:31.70&-15:39: 8.00&2004-04-18& 28.5&1.16&1.26&0.96&0.94&
\nodata&\nodata\\
IRAS20087-03&20:11:23.86& -2:59:50.80&2004-05-14& 49.1&0.99&0.96&0.87&0.86&
\nodata&0.62\\
IRAS20100-41&20:13:29.85&-41:47:34.70&2004-04-13& 44.1&0.89&0.90&0.86&0.82&
0.70&0.70\\
IRAS20414-16&20:44:18.18&-16:40:16.40&2004-05-14& 44.1&1.06&1.07&0.88&0.83&
\nodata&0.55\\
IRAS20551-42&20:58:26.78&-42:39: 1.60&2004-05-14& 39.7&1.07&1.11&0.97&0.92&
\nodata&0.85\\
IRAS21272+25&21:29:29.40& 25:27:55.10&2004-06-24& 30.4&1.16&1.18&1.12&1.13&
\nodata&\nodata\\
%IRAS21293-01&21:31:53.49& -1:41:43.40&2004-06-07& 72.5&****&1.21&0.20&1.19&
%\nodata&0.04\\
IRAS22491-18&22:51:49.35&-17:52:24.00&2004-06-24& 44.2&1.37&1.38&1.11&1.11&
\nodata&0.99\\
IRAS23128-59&23:15:47.01&-59: 3:16.90&2004-05-11& 42.4&1.59&1.73&1.43&1.38&
\nodata&1.11\\
IRAS23129+25&23:15:21.41& 26: 4:32.60&2003-12-17& 40.6&0.92&0.92&0.92&0.91&
\nodata&\nodata\\
IRAS23365+36&23:39: 1.29& 36:21: 9.80&2004-01-08& 43.1&2.90&2.68&1.31&1.17&
2.33&1.08\\
IRAS23498+24&23:52:26.05& 24:40:16.20&2004-01-04& 64.1&0.97&1.03&0.81&0.84&
0.68&0.53\\
IRAS23529-21&23:55:33.22&-21: 2:59.60&2003-12-16& 40.6&1.00&0.99&0.87&0.82&
\nodata&\nodata\\
Mrk1014& 1:59:50.23&  0:23:40.50&2004-01-07& 43.0&1.17&1.32&1.09&1.04&1.08&
0.86\\
Mrk231&12:56:14.29& 56:52:25.10&2004-04-14& 41.0&1.20&1.17&1.03&1.03&1.18&1.09
\\
Mrk273&13:44:42.12& 55:53:13.10&2004-04-14& 39.8&1.18&1.19&1.06&1.06&1.24&1.07
\\
Mrk463E&13:56: 2.90& 18:22:19.00&2004-01-07& 42.5&1.00&1.02&0.97&1.01&0.99&
0.97\\
NGC6240&16:52:58.89&  2:24: 3.40&2004-03-04& 40.0&0.93&1.08&1.00&1.01&1.10&
1.03\\
UGC5101& 9:35:51.65& 61:21:11.30&2004-03-23& 42.7&1.11&1.19&1.13&1.15&1.19&
1.10\\
\enddata
\end{deluxetable}
\clearpage

%\clearpage

%
% table 3 part a
%
%
\begin{deluxetable}{lcccccc} 
\tabletypesize{\scriptsize}
\tablecolumns{10} 
\tablewidth{0pc}
\tablecaption{Measured Line Fluxes and Limits}
\tablehead{
\colhead{Object } & \colhead{S(0)~28.22 $\mu m$} & \colhead{S(1)~17.04 $\mu m$} & 
\colhead{S(2)~12.28 $\mu m$} & \colhead{S(3)~9.67 $\mu m$ } & \colhead{S(7)~5.51 $\mu m$ } &  
\colhead{L$_{\rm H_2}$/L$_{\rm IR}$\tablenotemark{a}} \\
\colhead{} & \colhead{ (10$^{-20}$ W~cm$^{-2}$) } & \colhead{(10$^{-20}$ W~cm$^{-2}$) } & \colhead{ (10$^{-20}$ W~cm$^{-2}$) } &
\colhead{ (10$^{-20}$ W~cm$^{-2}$) } & \colhead{ (10$^{-20}$ W~cm$^{-2}$) } & \colhead{( $\%$ )} }
\startdata
3C273           & $<$~2.2 	& $<$3.5 	  	& $<$3.2 		& $<$3.6 		& $<$27.3		& \nodata\\
Arp220          & $<$97.0 	& 1.862$\pm$0.168  	& 0.98$\pm$0.13 	& 0.73$\pm$0.02\tablenotemark{b} 	& 1.29$\pm$0.38 	& 0.006\\
IRAS~00188-0856 & $<$~1.9 	& 0.043$\pm$0.005 	& 0.020$\pm$0.004 	& 0.038$\pm$0.012 	& $<$3.2 		& 0.004\\
IRAS~00275-0044 & $<$~3.7\tablenotemark{b}	& $<$1.2\tablenotemark{b}  	& $<$1.1\tablenotemark{b} 		& $<$0.5\tablenotemark{b} 		& $<$0.8 		& \nodata\\
IRAS~00275-2859 & $<$~5.4\tablenotemark{b}	& $<$3.1\tablenotemark{b}  	& $<$2.6\tablenotemark{b} 		& $<$2.7\tablenotemark{b} 		& $<$3.0 		& \nodata\\
IRAS~00397-1312 & $<$~1.6 	& 0.029$\pm$0.004 	& 0.017$\pm$0.002 	& 0.023$\pm$0.006 	& $<$8.0 		& 0.006\\
IRAS~00406-3127 & $<$~5.1\tablenotemark{b} & $<$1.5\tablenotemark{b}  	& $<$1.2\tablenotemark{b} 		& 0.019$\pm$0.001\tablenotemark{b} & $<$1.0 		& 0.003\\
IRAS~00476-0054 & \nodata 	& $<$0.1 	  	& $<$0.4\tablenotemark{b} 		& $<$0.3\tablenotemark{b} 		& $<$0.2 		& \nodata\\
IRAS~01003-2238 & $<$~3.5 	& 0.087$\pm$0.031 	& 0.034$\pm$0.002 	& 0.057$\pm$0.007 	& $<$2.5 		& 0.008\\
IRAS~01199-2307 & $<$10.6\tablenotemark{b} & $<$1.5\tablenotemark{b}  	& $<$1.2\tablenotemark{b} 		& $<$0.1\tablenotemark{b} 		& $<$0.5 		& \nodata\\
IRAS~02115+0226 & \nodata 	& $<$0.2\tablenotemark{b}  	& $<$0.2\tablenotemark{b} 		& 0.029$\pm$0.002\tablenotemark{b} & $<$0.03 		& 0.004\\
IRAS~02433+0110 & \nodata 	& $<$1.0\tablenotemark{b}  	& $<$0.8\tablenotemark{b} 		& $<$0.6\tablenotemark{b} 		& $<$0.4 		& \nodata\\
IRAS~03521+0028 & $<$~1.3 	& 0.082$\pm$0.013 	& 0.027$\pm$0.009 	& 0.037$\pm$0.003 	& $<$0.3 		& 0.006\\
IRAS~04114-5117 & $<$~8.7\tablenotemark{b} & 0.104$\pm$0.010\tablenotemark{b} & $<$0.6\tablenotemark{b} 		& 0.051$\pm$0.019\tablenotemark{b} & $<$0.4 		& 0.009\\
IRAS~04313-1649 & $<$~6.6\tablenotemark{b} & $<$0.9\tablenotemark{b} 		& $<$0.5\tablenotemark{b} 		& $<$0.1\tablenotemark{b} 		& $<$0.2 		& \nodata\\
IRAS~05189-2524 & $<$28.2 	& 0.34$\pm$0.07 		&   0.15$\pm$0.02		& 0.36$\pm$0.13 	& $<$34.1 		& 0.003\\
IRAS~06035-7102 & $<$~6.1 	& 0.417$\pm$0.004 	& 0.233$\pm$0.058 	& 0.34$\pm$0.10 	& $<$11.6 		& 0.023\\
IRAS~06206-6315 & $<$~2.9 	& 0.129$\pm$0.010 	& 0.050$\pm$0.005 	& 0.059$\pm$0.019 	& $<$1.8 		& 0.008\\
IRAS~06301-7934 & $<$~9.6\tablenotemark{b} & 0.107$\pm$0.020\tablenotemark{b} & 0.044$\pm$0.016\tablenotemark{b} & 0.072$\pm$0.017\tablenotemark{b} & 0.095$\pm$0.018 	& 0.022\\
IRAS~06361-6217 & $<$14.0\tablenotemark{b} & 0.094$\pm$0.004\tablenotemark{b} & $<$2.1\tablenotemark{b}  	& $<$0.3\tablenotemark{b} 		& $<$2.3 		& 0.011\\
IRAS~07598+6508 & $<$~3.0 	& 0.138$\pm$0.070 	& 0.037$\pm$0.006 	& 0.067$\pm$0.002 	& $<$22.5 		& 0.011\\
IRAS~08311-2459 & $<$~7.5 	& 0.388$\pm$0.030 	& 0.252$\pm$0.026 	& 0.32$\pm$0.10 	& $<$2.6 		& 0.019\\
IRAS~08572+3915 & $<$15.4 	& 0.119$\pm$0.004 	& 0.051$\pm$0.010 		& 0.046$\pm$0.013 	& $<$43.0 		& 0.003\\
IRAS~09022-3615\tablenotemark{c} & $<$~3.3 	& 0.122$\pm$0.008 	& $<$0.2 		& $<$0.2 		& $<$0.04 		& 0.001\\
IRAS~09463+8141 & $<$~4.4\tablenotemark{b} & 0.108$\pm$0.055\tablenotemark{b} & 0.045$\pm$0.022\tablenotemark{b} & 0.083$\pm$0.007\tablenotemark{b} & $<$0.4 		& 0.019\\
IRAS~10091+4704 & $<$~4.5\tablenotemark{b} & $<$0.4\tablenotemark{b} 		& $<$0.4\tablenotemark{b}  	& $<$0.1\tablenotemark{b} 		& $<$0.3 		& \nodata\\
IRAS~10398+3247 & \nodata 	& $<$0.1\tablenotemark{b} 		& $<$0.001\tablenotemark{b} 	& $<$0.004\tablenotemark{b} 	& $<$0.2 		& \nodata\\
IRAS~10565+2448 & $<$10.9 	& 0.657$\pm$0.005 	& 0.256$\pm$0.009 	& 0.403$\pm$0.002 	& $<$5.4 		& 0.013\\
IRAS~11119+3257 & $<$~2.5 	& $<$1.7   		& $<$1.6 		& 0.054$\pm$0.007 	& $<$9.1 		& 0.003\\
IRAS~12018+1941 & $<$~2.6 	& 0.093$\pm$0.006 	& 0.030$\pm$0.003 	& 0.043$\pm$0.008 	& $<$1.1 		& 0.010\\
IRAS~12032+1707 & $<$10.7\tablenotemark{b} & 0.094$\pm$0.030\tablenotemark{b} & $<$1.5\tablenotemark{b} 		& 0.077$\pm$0.027\tablenotemark{b} & 0.147$\pm$0.044 	& 0.028\\
IRAS~12072-0444 & $<$~4.5 	& 0.192$\pm$0.017 	& 0.089$\pm$0.010 	& 0.136$\pm$0.008 	& $<$2.7 		& 0.020\\
IRAS~12112+0305 & 0.14$\pm$0.04 & 0.412$\pm$0.044 	& 0.173$\pm$0.031 	& 0.237$\pm$0.003 	& 0.41$\pm$0.08 	& 0.021\\
IRAS~12514+1027 & $<$~0.7 	& $<$0.5  		& $<$0.5 		& $<$0.2 		& $<$4.0 		& \nodata\\
IRAS~13120-5453 & $<$30.9 	& 1.047$\pm$0.034 	& 0.662$\pm$0.045 	& 0.763$\pm$0.005 	& $<$14.6 		& 0.008\\
IRAS~13218+0552 & $<$~1.8 	& $<$1.5   		& $<$8.9\tablenotemark{b} 	  	& $<$9.4\tablenotemark{b} 		& $<$16.9 		& \nodata\\
IRAS~13342+3932 & 0.03$\pm$0.01 & 0.102$\pm$0.002 	& $<$3.9\tablenotemark{b} 		& $<$4.2\tablenotemark{b} 		& $<$2.6 		& 0.016\\
IRAS~13352+6402 & $<$~5.2\tablenotemark{b} & $<$1.1\tablenotemark{b}  	& $<$0.9\tablenotemark{b} 		& 0.016$\pm$0.008\tablenotemark{b} & $<$1.2 		& 0.002\\
IRAS~13451+1232 & $<$~4.5 	& 0.283$\pm$0.065 	& 0.129$\pm$0.005 	& 0.219$\pm$0.026 	& $<$4.1 		& 0.031\\
IRAS~14348-1447 & $<$~6.1 	& 0.447$\pm$0.015 	& 0.195$\pm$0.010 	& 0.240$\pm$0.021 	& $<$3.1 		& 0.017\\
IRAS~14378-3651 & $<$~6.6 	& 0.193$\pm$0.016 	& 0.079$\pm$0.016 	& 0.125$\pm$0.180 	& $<$2.3 		& 0.008\\
IRAS~14537+1950 & \nodata 	& $<$0.1   		& $<$0.2\tablenotemark{b} 		& $<$0.1\tablenotemark{b} 		& $<$0.1 		& \nodata\\
IRAS~14548+3349 & \nodata 	& $<$0.2\tablenotemark{b} 	        & $<$0.2\tablenotemark{b} 		& $<$0.1\tablenotemark{b} 		& $<$0.2 		& \nodata\\
IRAS~15001+1433 & $<$~1.9 	& 0.077$\pm$0.007 	& $<$2.8\tablenotemark{b} 		& 0.068$\pm$0.013\tablenotemark{b} & $<$1.7 		& 0.010\\
IRAS~15206+3342 & $<$~2.5       & 0.102$\pm$0.053\tablenotemark{b} & $<$4.3\tablenotemark{b} 		& $<$2.6\tablenotemark{b} 		& $<$1.6 		& 0.006\\
IRAS~15250+3609 & $<$12.9 	& 0.156$\pm$0.014 	& 0.050$\pm$0.006 	& 0.048$\pm$0.004 	& $<$10.4 		& 0.004\\
IRAS~15462-0450 & $<$~4.1 	& 0.134$\pm$0.006 	& 0.037$\pm$0.001 	& 0.086$\pm$0.012 	& $<$5.7 		& 0.010\\
IRAS~16124+3241 & \nodata 	& $<$0.3\tablenotemark{b}   	& $<$0.1\tablenotemark{b} 		& $<$0.1\tablenotemark{b} 		& $<$0.1 		& \nodata\\
IRAS~16334+4630 & $<$~4.2\tablenotemark{b} & $<$0.9\tablenotemark{b}    	& $<$0.7\tablenotemark{b} 		& $<$0.4\tablenotemark{b} 		& $<$0.4 		& \nodata\\
IRAS~17068+4027 & $<$10.2\tablenotemark{b} & 0.072$\pm$0.018\tablenotemark{b} & $<$1.7\tablenotemark{b} 		& 0.052$\pm$0.017\tablenotemark{b} & 0.087$\pm$0.019 	& 0.020\\
IRAS~17179+5444 & $<$~1.7 	& 0.209$\pm$0.007 	& 0.068$\pm$0.024 	& 0.14$\pm$0.02 	& $<$1.4 		& 0.031\\
IRAS~17208-0014 & $<$23.9 	& 0.881$\pm$0.009 	& 0.497$\pm$0.085 	& 0.57$\pm$0.11 	& $<$8.5 		& 0.009\\
IRAS~17233+3712 & \nodata 	& $<$0.1    		& $<$0.5\tablenotemark{b} 		& $<$0.3\tablenotemark{b} 		& $<$0.1 		& \nodata\\
IRAS~17463+5806 & $<$~3.5\tablenotemark{b} & $<$0.6\tablenotemark{b} 	        & $<$0.3\tablenotemark{b} 		& 0.023$\pm$0.011\tablenotemark{b} & $<$0.2 		& 0.007\\
IRAS~18030+0705 & $<$~1.4\tablenotemark{b} & 0.111$\pm$0.008\tablenotemark{b} & $<$0.8\tablenotemark{b} 		& $<$0.4\tablenotemark{b} 		& $<$0.2 		& 0.008\\
IRAS~18443+7433 & $<$15.5\tablenotemark{b} & $<$2.7\tablenotemark{b} 	        & $<$2.4\tablenotemark{b} 		& 0.048$\pm$0.007\tablenotemark{b} & $<$1.7 		& 0.003\\
IRAS~19254-7245 & $<$~8.5 	& 0.881$\pm$0.057 	& 0.382$\pm$0.095 	& 0.382$\pm$0.004 	& $<$0.4 		& 0.031\\
IRAS~19297-0406 & $<$~4.8 	& 0.332$\pm$0.082\tablenotemark{b} & $<$3.3\tablenotemark{b} 		& 0.196$\pm$0.008\tablenotemark{b} & $<$1.7 		& 0.009\\
IRAS~19458+0944 & $<$19.3\tablenotemark{b} & 0.219$\pm$0.042\tablenotemark{b} & $<$2.2\tablenotemark{b} 		& 0.113$\pm$0.038\tablenotemark{b} & $<$1.0 		& 0.009\\
IRAS~20037-1547 & $<$11.2\tablenotemark{b} & $<$5.8\tablenotemark{b} 		& $<$4.4\tablenotemark{b} 		& 0.081$\pm$0.005\tablenotemark{b} & $<$4.8 		& 0.006\\
IRAS~20087-0308 & $<$~2.1 	& 0.229$\pm$0.014\tablenotemark{b} & 0.084$\pm$0.033\tablenotemark{b} & 0.13$\pm$0.04\tablenotemark{b} 	& $<$2.2 		& 0.011\\
IRAS~20100-4156 & $<$~4.6 	& 0.083$\pm$0.007 	& 0.032$\pm$0.001 	& 0.068$\pm$0.010 	& $<$2.9 		& 0.005\\
IRAS~20414-1651 & $<$~2.4 	& 0.156$\pm$0.003\tablenotemark{b} & $<$1.6\tablenotemark{b} 		& 0.049$\pm$0.033\tablenotemark{b} & $<$0.7	 	& 0.006\\
IRAS~20551-4250 & $<$16.3 	& 0.896$\pm$0.052\tablenotemark{b} & $<$18.0\tablenotemark{b} 	& 0.443$\pm$0.020\tablenotemark{b} & $<$9.8 		& 0.013\\
IRAS~21272+2514 & $<$~6.6\tablenotemark{b} & $<$1.0\tablenotemark{b} 		& $<$0.7\tablenotemark{b} 		& 0.037$\pm$0.002\tablenotemark{b} & $<$0.5 		& 0.004\\
IRAS~22491-1808 & $<$~7.0 	& 0.165$\pm$0.005\tablenotemark{b} & $<$3.1\tablenotemark{b} 		& 0.098$\pm$0.013\tablenotemark{b} & $<$1.0 		& 0.006\\
IRAS~23128-5919 & $<$12.2 	& 0.514$\pm$0.010\tablenotemark{b} & $<$22.595\tablenotemark{b} 	& 0.305$\pm$0.080\tablenotemark{b} & $<$7.3 		& 0.009\\
IRAS~23129+2548 & $<$10.1\tablenotemark{b} & $<$1.3\tablenotemark{b} 		& $<$1.1\tablenotemark{b} 		& $<$0.1\tablenotemark{b} 		& $<$0.6 		& \nodata\\
IRAS~23365+3604 & $<$~9.1 	& 0.390$\pm$0.031 	& 0.201$\pm$0.095 	& 0.31$\pm$0.05 	& $<$3.0 		& 0.015\\
IRAS~23498+2423 & $<$0.68 	& $<$0.5 		& $<$0.3    		& $<$0.2 		& $<$2.1 		& \nodata\\
IRAS~23529-2119 & \nodata 	& $<$1.4\tablenotemark{b} 		& $<$1.0\tablenotemark{b} 		& 0.028$\pm$0.011\tablenotemark{b} & $<$0.7 		& 0.010\\
Mrk~231\tablenotemark{d}         & $<$67.5 	& $<$31.7 		& $<$26.8 		& 0.46$\pm$0.13 	& $<$122.3 		& 0.001\\
Mrk~273         & $<$26.25 	& 1.024$\pm$0.009 	& 0.56$\pm$0.07 	& 1.04$\pm$0.09 	& $<$15.0 		& 0.015\\
Mrk~463E        & $<$~7.9 	& 0.279$\pm$0.037 	& 0.131$\pm$0.008 	& 0.28$\pm$0.06 	& $<$38.5 		& 0.017\\
Mrk~1014        & $<$~4.2 	& $<$2.5 		& $<$1.5 		& 0.07$\pm$0.03 	& $<$4.6 		& 0.003\\
NGC~6240        & 0.50$\pm$0.25 & 4.271$\pm$0.344 	& 3.55$\pm$0.26 	& 5.83$\pm$0.43 	& 5.54$\pm$0.50 	& 0.097\\
UGC~5101        & $<$10.0 	& 0.496$\pm$0.047 	& 0.27$\pm$0.05 	& 0.28$\pm$0.03 	& $<$19.5 		& 0.010\\
\enddata
\tablenotetext{a} {Ratio of the sum of the H$_{\rm 2}$ line fluxes divided by the IR luminosity 
in \%. The far-infrared luminosities are defined in Table 2.}

\tablenotetext{b} {Emission lines observed using IRS-LORES.}
 
\tablenotetext{c} {The absolute flux calibration for IRAS~09022-3615 is uncertain, see text}

\tablenotetext{d} {Both the S(1) and S(2) lines are present. However
  we quote upper limits as both lines are poorly calibrated. The S(1)
  line straddles orders 12 and 11 in IRS-SH and the centre of S(2) line profile is
  affected by bad pixel values.}
\end{deluxetable}
\clearpage

%\clearpage

%
% table 4 
%
%
\begin{deluxetable}{lcccc}
\tablecolumns{5} 
\tablewidth{0pc}
\tablecaption{Properties of the Warm Molecular \mh }
\tablehead{
\colhead{Object}&
\colhead{T$_{\rm ex}$}&
\colhead{\rm Warm M$_{\rm H_2}$ }&
\colhead{Cold M$_{\rm H_2}$}&
\colhead{ $\frac{\rm Warm~M_{\rm H_2}}{\rm Cold~M_{\rm H_2}}$       }\\
\colhead{}&
\colhead{( K )}&
\colhead{(10$^7$~M$_{\odot}$)}&
\colhead{(10$^{10}$~M$_{\odot}$)}&
\colhead{( \% )}
}
\startdata
Arp 220&1435$\pm$20&  0.03 $\pm$0.02  & & \\
&258$\pm$4&  6.6 $\pm$0.8  &  & \\
&&&3.72$^{R1}$&0.18\\
IRAS 00188-0856&358$\pm$25&   4.11$\pm$   1.18&   3.98$^{R1}$& 0.10\\
IRAS 00397-1312&365$\pm$24&  13.16$\pm$   3.69&\nodata&\nodata\\
IRAS~00406-3127$^L$&$<336\pm15>$&  32.04$\pm$
  11.95&\nodata&\nodata\\
IRAS~01003-2238&365$\pm$19&   6.68$\pm$   2.71&\nodata&\nodata\\
IRAS~02115+0226$^L$&$<336\pm15>$&  69.14$\pm$
  26.08&\nodata&\nodata\\
IRAS~03521+0028&317$\pm$11&  14.67$\pm$   3.14&   5.62$^{R1}$& 0.26\\
IRAS~04114-5117$^L$&321$\pm$26&  11.74$\pm$
   4.05&\nodata&\nodata\\
IRAS~05189-2524&380$\pm$30& 2.86$\pm$1.82&   0.26$^{R2}$& 1.1\\
IRAS~06035-7102&363$\pm$23&  13.94$\pm$   3.30&   4.17$^{R3}$& 0.33\\
IRAS~06206-6315&313$\pm$15&   8.14$\pm$   1.76&   9.33$^{R3}$& 0.09\\
IRAS~06301-7934$^L$&1329$\pm$106&  0.3$\pm$   0.1&\nodata&\nodata\\
&264$\pm$14&  32$\pm$ 10&\nodata&\nodata\\
IRAS~06361-6217$^L$&$<336\pm15>$&  16.56$\pm$
   2.93&\nodata&\nodata\\
IRAS~07598+6508&366$\pm$22&  17.53$\pm$   9.65&   6.76$^{R1}$& 0.26\\
IRAS~08311-2459&384$\pm$22&  19.27$\pm$   4.23&\nodata&\nodata\\
IRAS~08572+3915&310$\pm$18&   2.93$\pm$   0.76&   0.72$^{R3}$&
 0.41\\
IRAS~09022-3615&$<336\pm15>$&   2.61$\pm$   0.48&\nodata&
\nodata\\
IRAS~09463+8141$^L$&363$\pm$39&  15.10$\pm$   9.80&\nodata&\nodata
\\
IRAS~10565+2448&338$\pm$1&   7.05$\pm$   0.10&   3.02$^{R3}$& 0.23\\
IRAS~11119+3257&$<336\pm15>$&  23.21$\pm$   9.14&\nodata&
\nodata\\
IRAS~12018+1941&309$\pm$11&  22.26$\pm$   3.63&\nodata&\nodata\\
IRAS~12032+1707$^L$&1471$\pm$196&  0.5$\pm$0.5&\nodata&\nodata\\
&309$\pm$30&  38$\pm$20&\nodata&\nodata\\
IRAS~12072-0444&349$\pm$8&  19.32$\pm$   2.39&\nodata&\nodata\\
IRAS~12112+0305&1428$\pm$1049&  0.2$\pm$0.8   &   & \\
&271$\pm$5& 119$\pm$41  &   & \\
&&&0.49$^{R4}$&24\\
IRAS~13120-5453&348$\pm$3&   5.29$\pm$   0.24&\nodata&\nodata\\
IRAS~13342+3932&141$\pm$15 &  803$\pm$459 (s0)&\nodata&\nodata\\
&$<336\pm15>$&  23.15$\pm$   0.46&\nodata&\nodata\\
IRAS~13352+6402$^L$&$<336\pm15>$&  11.10$\pm$
   7.04&\nodata&\nodata\\
IRAS~13451+1232&366$\pm$18&  23.30$\pm$   6.85&   0.51$^{R4}$& 4.57\\
IRAS~14348-1447&328$\pm$6&  19.86$\pm$   1.61&   3.47$^{R4}$& 0.57\\
IRAS~14378-3651&320$\pm$31&   5.97$\pm$   2.42&   2.29$^{R3}$& 0.26\\
IRAS~15001+1433$^L$&346$\pm$40&  17.05$\pm$
  11.10&\nodata&\nodata\\
IRAS~15206+3342$^L$&$<336\pm15>$&  10.45$\pm$
   5.72&\nodata&\nodata\\
IRAS~15250+3609&294$\pm$7&   3.89$\pm$   0.53&\nodata&\nodata\\
IRAS~15462-0450&285$\pm$5&  12.57$\pm$   1.15&\nodata&\nodata\\
IRAS~17068+4027$^L$&1416$\pm$154&  0.2$\pm$0.2&\nodata&\nodata\\
&292$\pm$23 &  22$\pm$9&\nodata&\nodata\\
IRAS~17179+5444&340$\pm$13&  30.03$\pm$   4.61&\nodata&\nodata\\
IRAS~17208-0014&347$\pm$13&   8.93$\pm$   1.29&   3.80$^{R1}$& 0.23\\
IRAS~17463+5806$^L$&$<336\pm15>$&  29.87$\pm$
  17.70&\nodata&\nodata\\
IRAS~18030+0705$^L$&$<336\pm15>$&  16.01$\pm$
   3.00&\nodata&\nodata\\
IRAS~18443+7433$^L$&$<336\pm15>$ &   9.71$\pm$
   3.87&\nodata&\nodata\\
IRAS~19254-7245&313$\pm$4&  23.63$\pm$   1.99&   4.57$^{R6}$& 0.52\\
IRAS~19297-0406$^L$&334$\pm$18 &  15.39$\pm$
   5.02&   5.25$^{R1}$& 0.29\\
IRAS~19458+0944$^L$&325$\pm$27 &  14.94$\pm$
   5.80&   6.76$^{R1}$& 0.22\\
IRAS~20037-1547$^L$&$<336\pm15>$&  35.92$\pm$
  13.50&\nodata&\nodata\\
IRAS~20087-0308$^L$&326$\pm$21&  17.49$\pm$   4.69&   9.33$^{R1}$& 0.1875
\\
IRAS~20100-4156&339$\pm$12&   9.05$\pm$   1.46&\nodata&\nodata\\
IRAS~20414-1651$^L$&294$\pm$38&  10.04$\pm$
   5.70&\nodata&\nodata\\
IRAS~20551-4250$^L$&322$\pm$5 &  10.46$\pm$
   0.90&   2.82$^{R6}$& 0.37\\
IRAS~21272+2514$^L$&$<336\pm15>$&   9.49$\pm$
   3.55&\nodata&\nodata\\
IRAS~22491-1808$^L$&335$\pm$10 &   6.09$\pm$
   0.74&   4.27$^{R3}$& 0.14\\
IRAS~23128-5919$^L$&335$\pm$19 &   6.04$\pm$
   1.36&   1.95$^{R3}$& 0.31\\
IRAS~23365+3604&358$\pm$14&   8.62$\pm$   1.45&   4.68$^{R1}$& 0.18\\
IRAS~23529-2119$^L$&$<336\pm15>$&  81.05$\pm$  43.67&\nodata&\nodata\\
Mrk 231&$<336\pm15>$&   7.97$\pm$   3.73&   4.07$^{R1}$& 0.20\\
Mrk 273&378$\pm$8&   6.75$\pm$   0.52&   2.69$^{R1}$& 0.25\\
Mrk 463E&368$\pm$21&   3.50$\pm$   0.87&   0.04$^{R4}$& 8.75\\
Mrk 1014&$<336\pm15>$&  20.38$\pm$  12.20&   5.25$^{R1}$& 0.39\\
NGC 6240&1327$\pm$36&  0.41$\pm$   0.08&  & \\
&292$\pm$6&  62$\pm$   32&  & \\
%&194$\pm$37 s0-s1 only&108$\pm$88&& \\
&&&4.37$^{R1}$&1.43\\
UGC 5101&332$\pm$11&   4.75$\pm$   0.78&   0.62$^{R2}$& 0.77\\
\enddata
%\tiny
\scriptsize \tablecomments{The Warm Molecular \mh Mass is derived from
  our data.  HIRES observations are used when available, and the LORES
  observations are marked in the first column with an `L'. The
  temperature is derived from the fit to the S(1) - S(3) data. The
  mass is derived using the S(1) line flux, if this line is not
  detected the S(3) line flux is used. If only a single line is
  detected we use the average temperature of 336$\pm$15 K to estimate
  the mass, denoted by `$<>$' in column 2.  The standard Milky Way \mh
  mass-to-CO luminosity ratio, $\alpha = $4.6 \msolar/K km s$^{-1}$
  pc$^{2}$ has been used to estimate the cold gas mass. The CO
  references labeled R1-R4 in column 3 are: R1: Solomon et al. 1997;
  R2: Rigopoulou et al. 1996; R3: Mirabel et al. 1990; R4: Evans et
  al. 2002. However, Solomon et al.  (1997) argue that $\alpha \sim$
  1.4 should be used for ULIRGs.  This would decrease the cold gas
  mass and correspondingly increase the warm gas fraction by a factor
  of $\sim$3. The S(7) line is detected in Arp
  220, IRAS 06301-7934, IRAS 12032+1707, IRAS 12112+0305, IRAS
  17068+4027 and NGC 6240, and we calculate a hot gas mass. The S(0)
  line is detected in IRAS 12112+0305, IRAS 13342+3932 and NGC 6240,
  which lowers the warm gas temperature. The S(0) line flux is
  used to estimate the gas mass.  For IRAS 13342+3932 we only detect
  the S(0) and S(1) line resulting in much lower gas temperature. We
  also quote the gas mass using the average temperature for the
  sample, which may be a more reasonable estimate of the gas mass.
  There is a problem with the absolute flux calibration for IRAS
  09022-3615 and the mass could be a factor twenty higher than
  quoted.}

\end{deluxetable}

\clearpage

\begin{figure}
\centering
%\subfigure[First Part]{
\label{fig:graphics:a}
\includegraphics[angle=0,scale=.7]{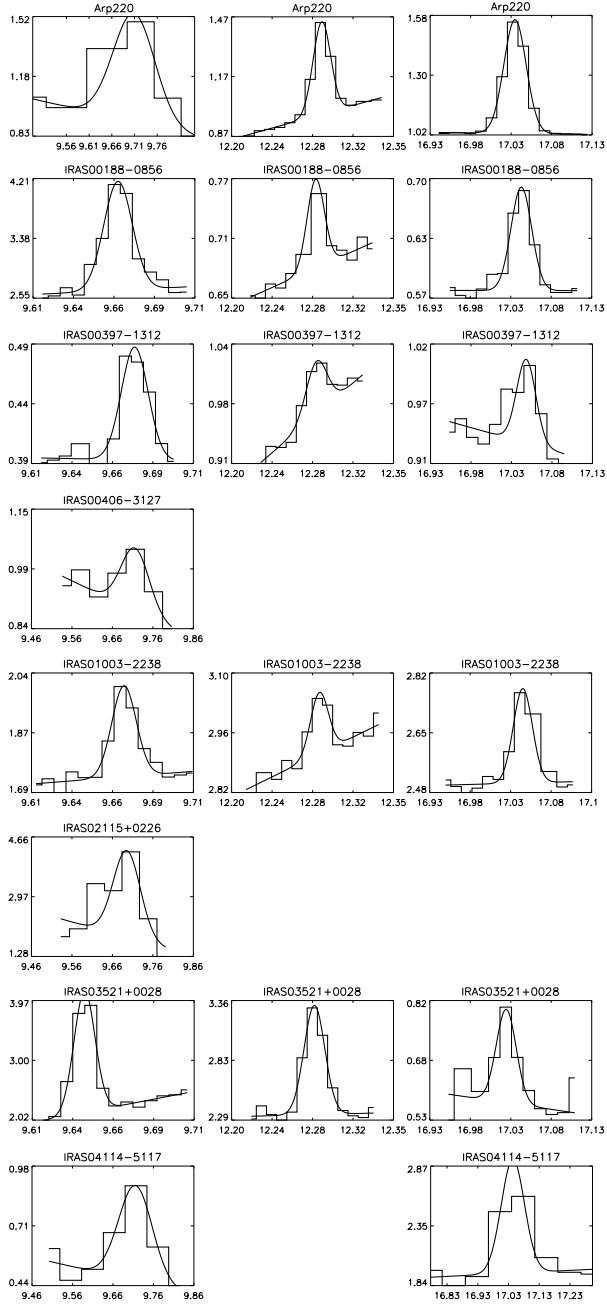} %}%
\caption{Pure rotational H$_{\rm 2}$ emission lines for the ULIRG
sample observed with Spitzer's IRS. The three columns show (from
left to right) the \mh S(3) 9.67 $\mu$m, S(2) 12.28 $\mu$m, and S(1)
17.04 $\mu$m lines after averaging over both slit positions. The
data are shown as a histogram, while the solid line shows a
single gaussian plus linear baseline fit. The vertical axis shows
flux density in units of 10$^{-18}$ W cm$^{-2}$ $\mu$m$^{-1}$, with
rest wavelength in microns shown on the horizontal axis.}
\label{fig:graphics}
\end{figure}

\addtocounter{figure}{-1}
\begin{figure}
\centering
\label{fig:graphics:b}
\includegraphics[angle=0,scale=.7]{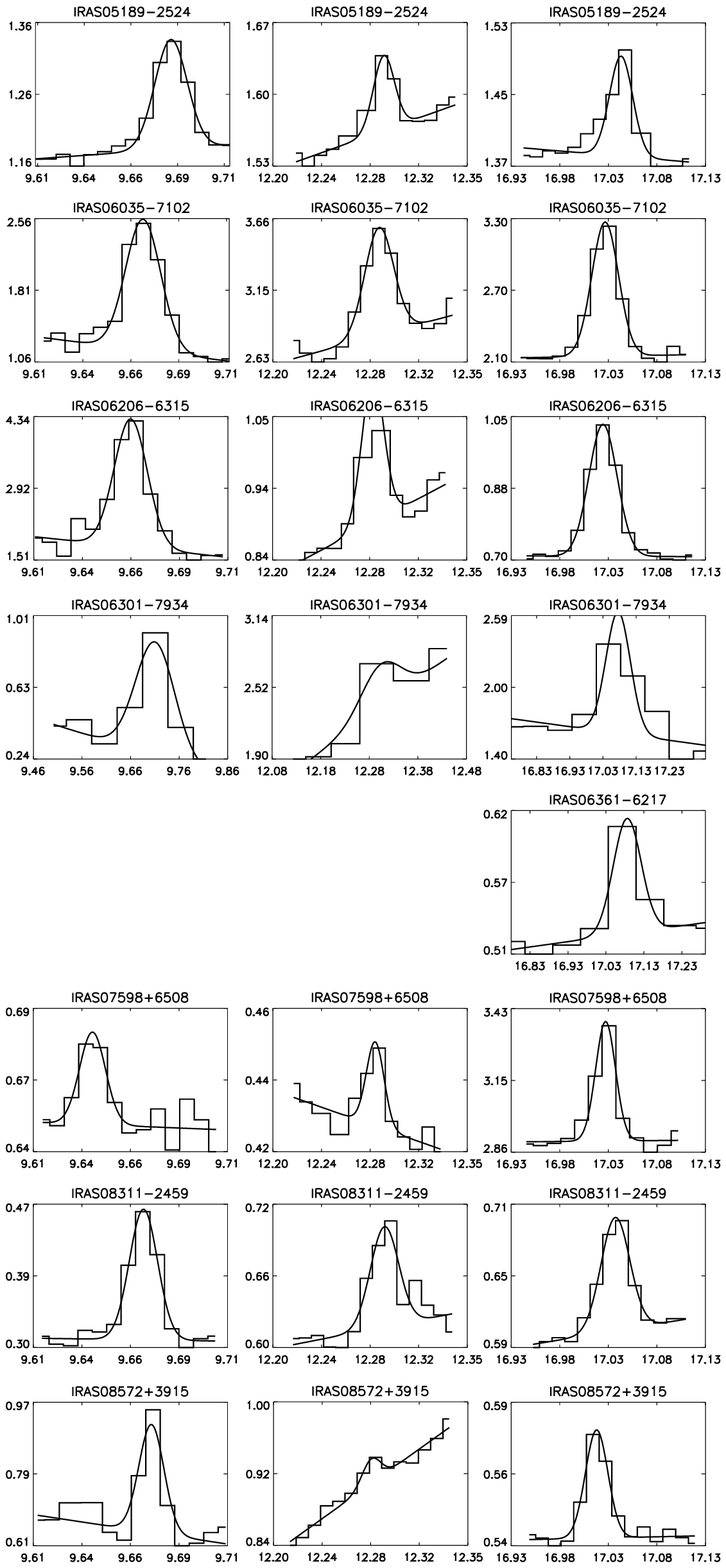} %}%
\caption{ Cont.}
\end{figure}

\addtocounter{figure}{-1}
\begin{figure}
\centering
\label{fig:graphics:b}
\includegraphics[angle=0,scale=.7]{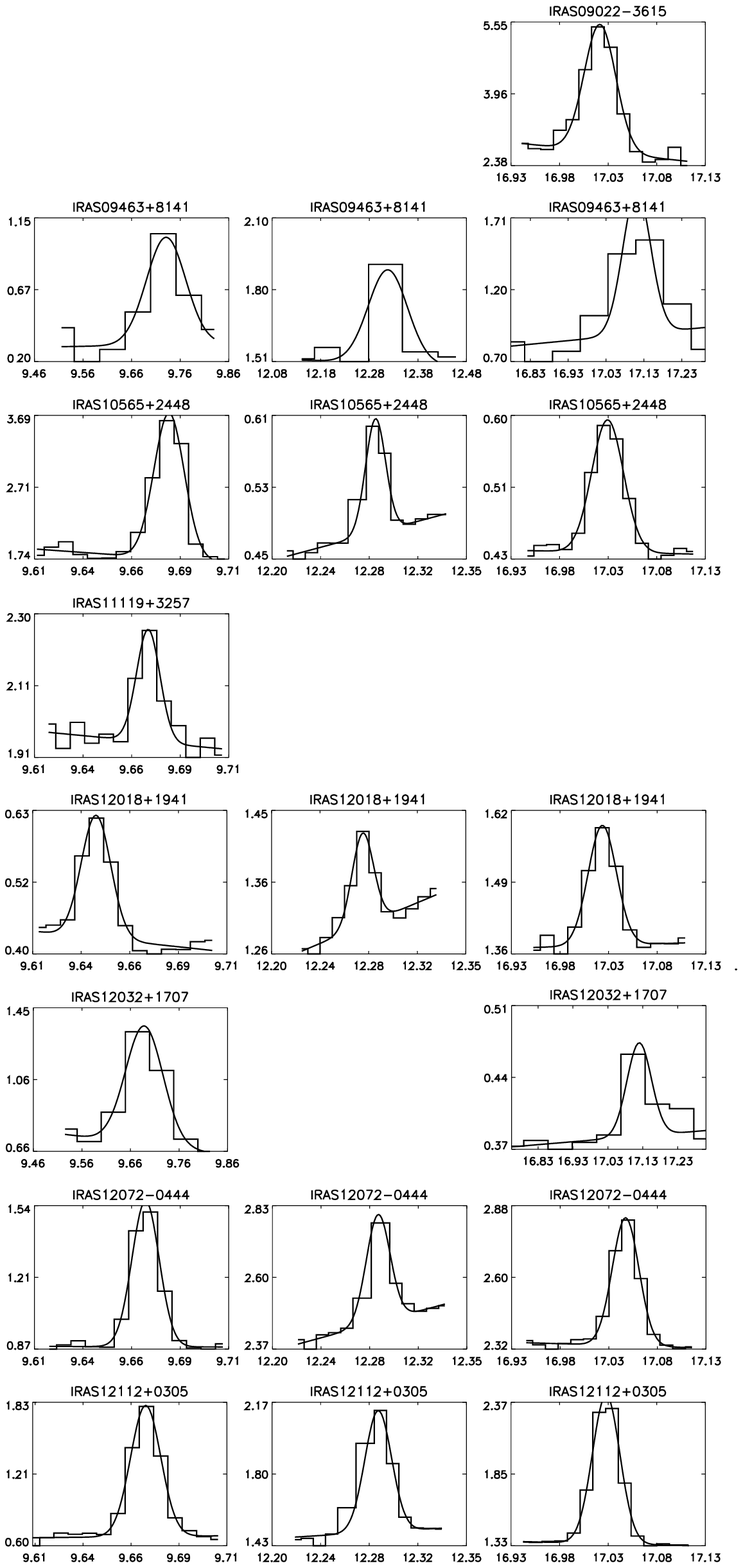} %}%
\caption{ Cont.}
\end{figure}

\addtocounter{figure}{-1}
\begin{figure}
\centering
\label{fig:graphics:b}
\includegraphics[angle=0,scale=.7]{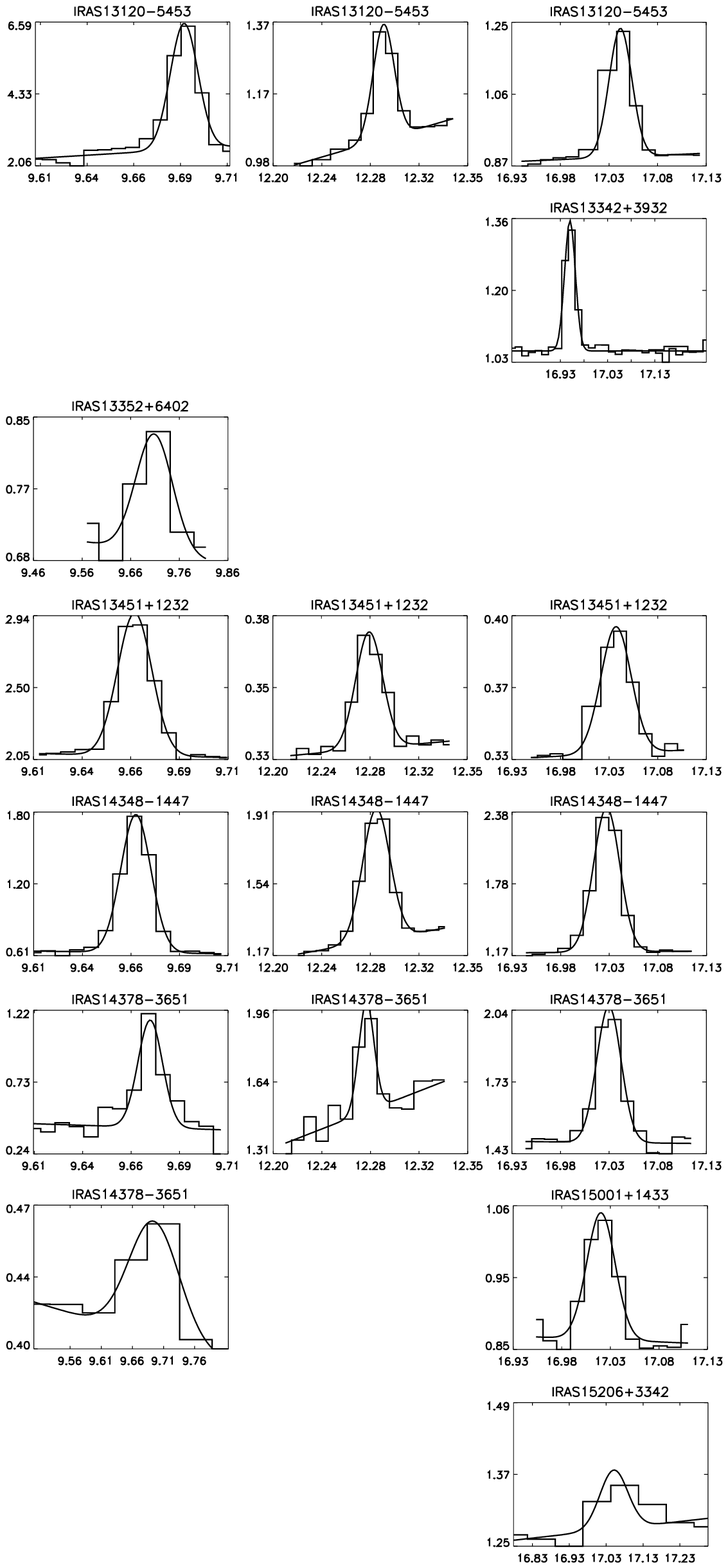} %}%
\caption{ Cont.}
\end{figure}

\addtocounter{figure}{-1}
\begin{figure}
\centering
\label{fig:graphics:b}
\includegraphics[angle=0,scale=.7]{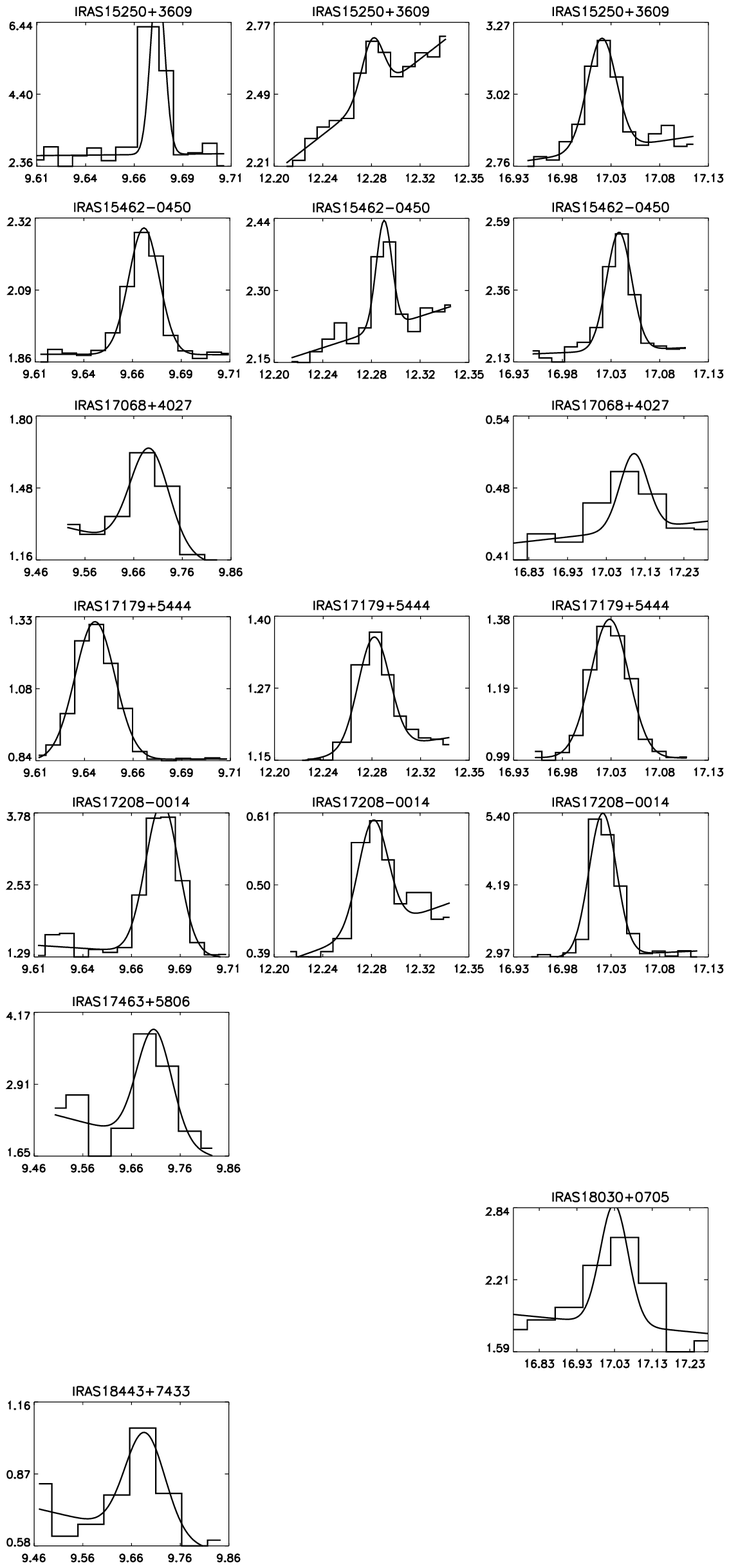} %}%
\caption{ Cont.}
\end{figure}

\addtocounter{figure}{-1}
\begin{figure}
\centering
\label{fig:graphics:b}
\includegraphics[angle=0,scale=.7]{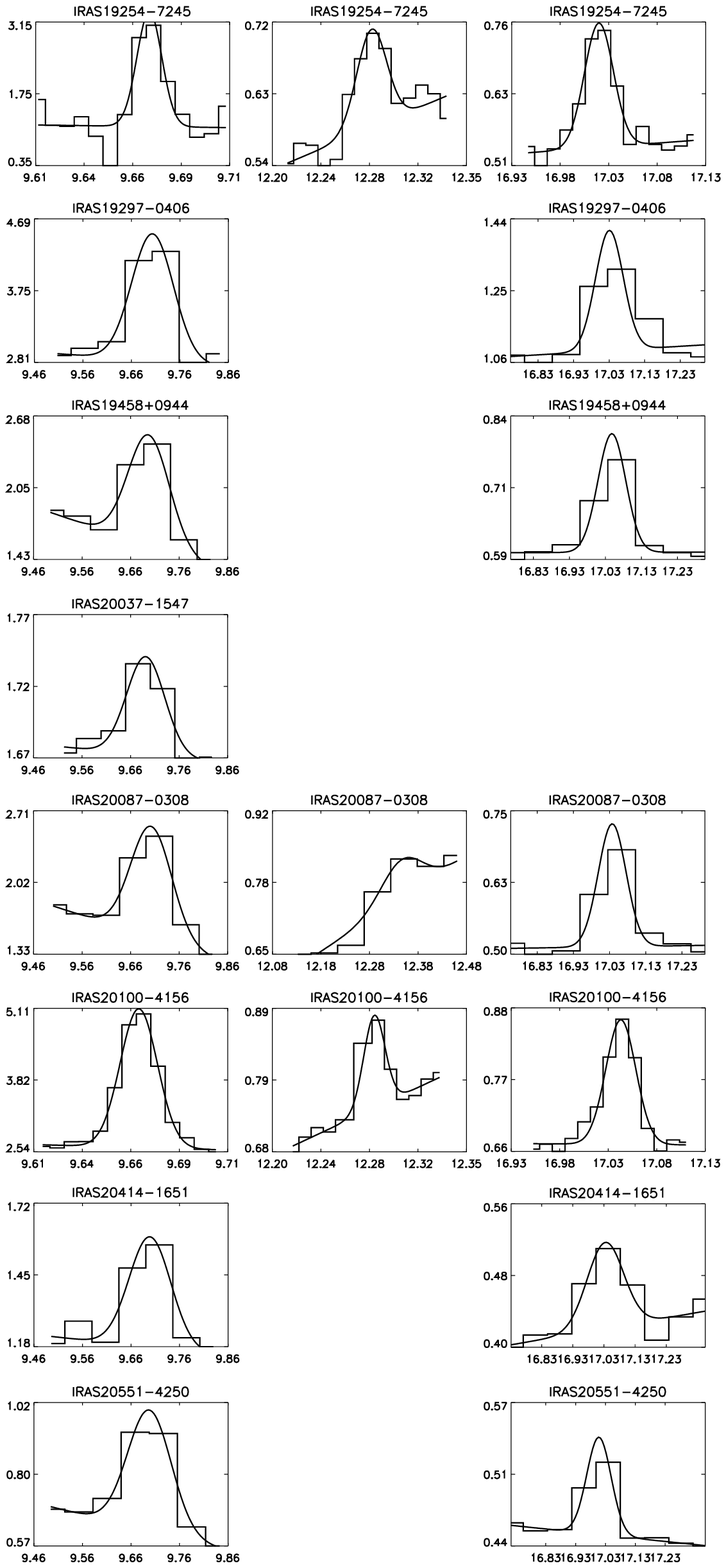} %}%
\caption{ Cont.}
\end{figure}

\addtocounter{figure}{-1}
\begin{figure}
\centering
\label{fig:graphics:b}
\includegraphics[angle=0,scale=.7]{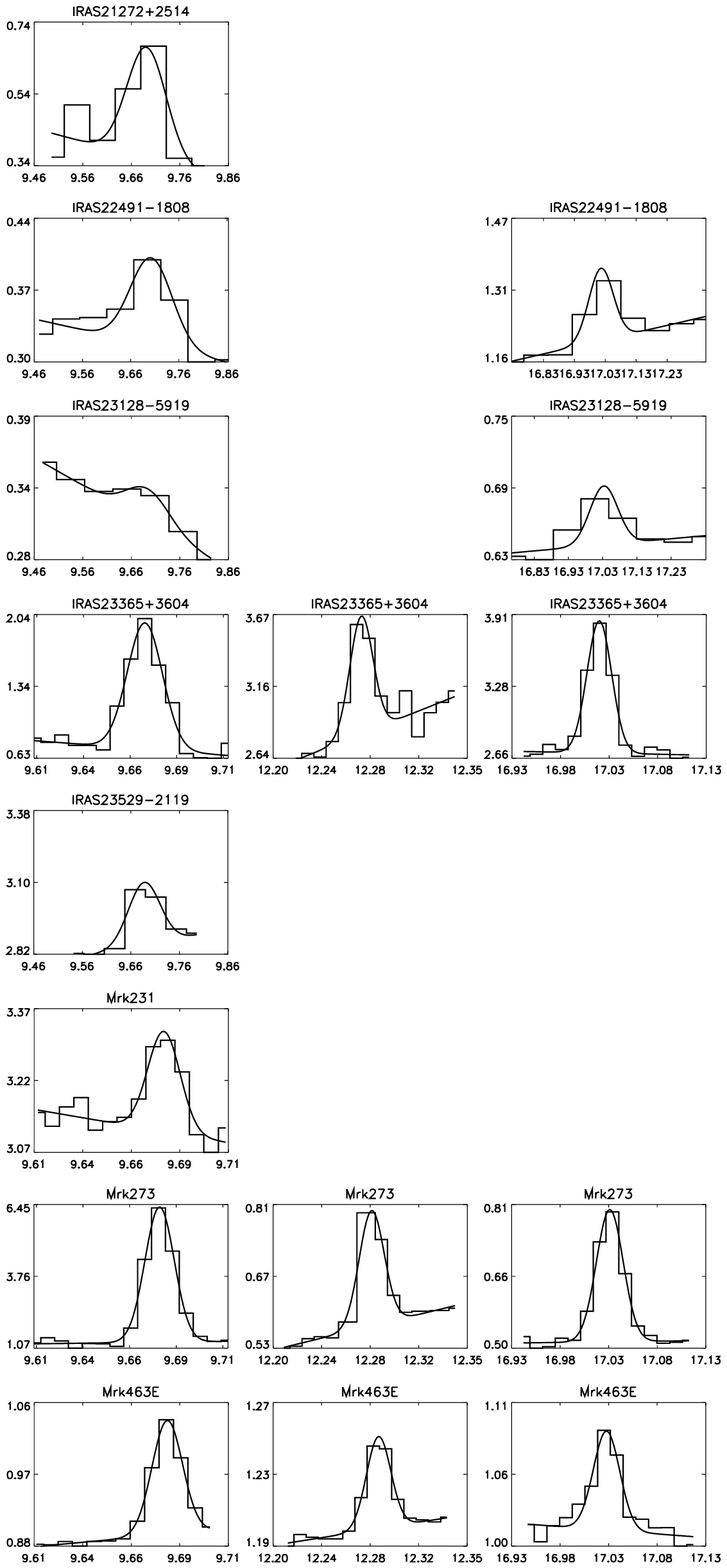} %}%
\caption{ Cont.}
\end{figure}

\addtocounter{figure}{-1}
\begin{figure}
\centering
\label{fig:graphics:b}
\includegraphics[angle=0,scale=.7]{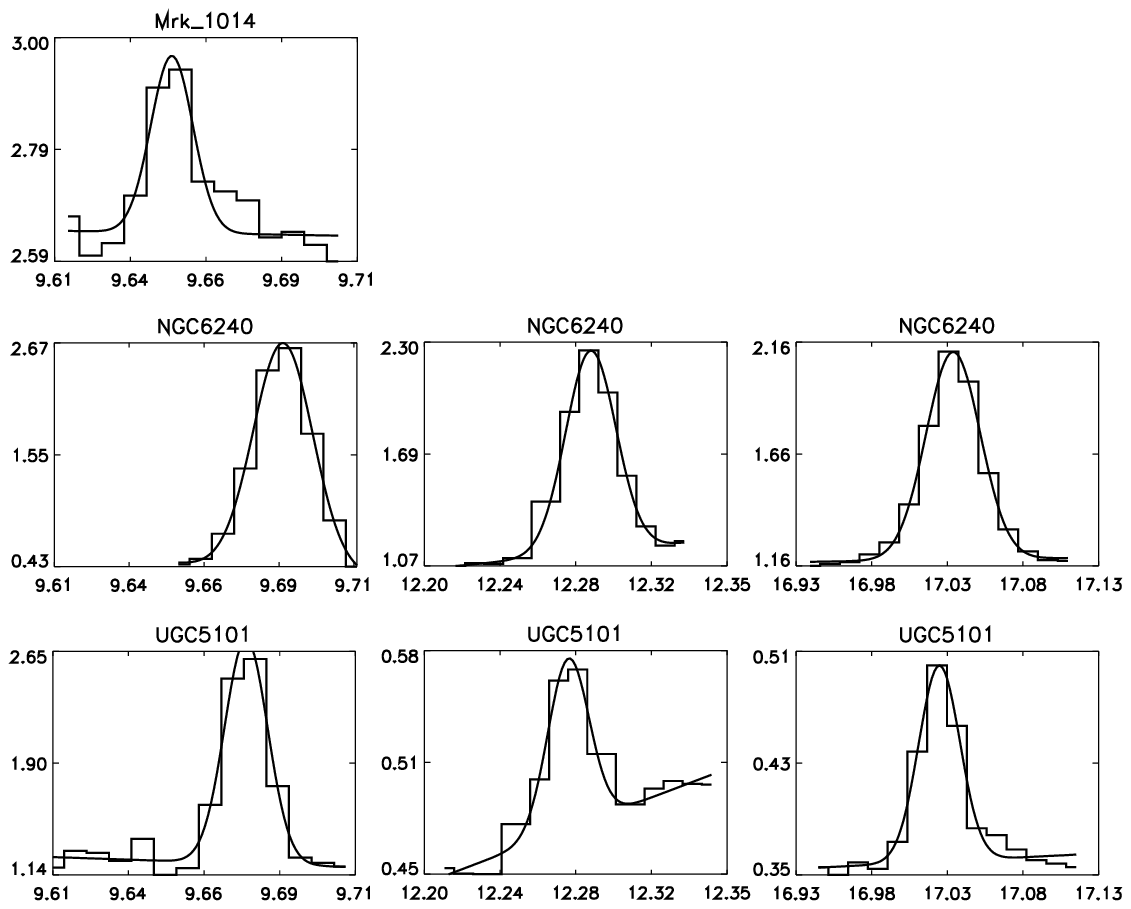} %}%
\caption{ Cont.}
\end{figure}

\clearpage

%FIGURE 2 Molecular Hydrogen s0
\begin{figure}
\centering
\label{fig:graphics:b}
\includegraphics[angle=0,scale=1.0]{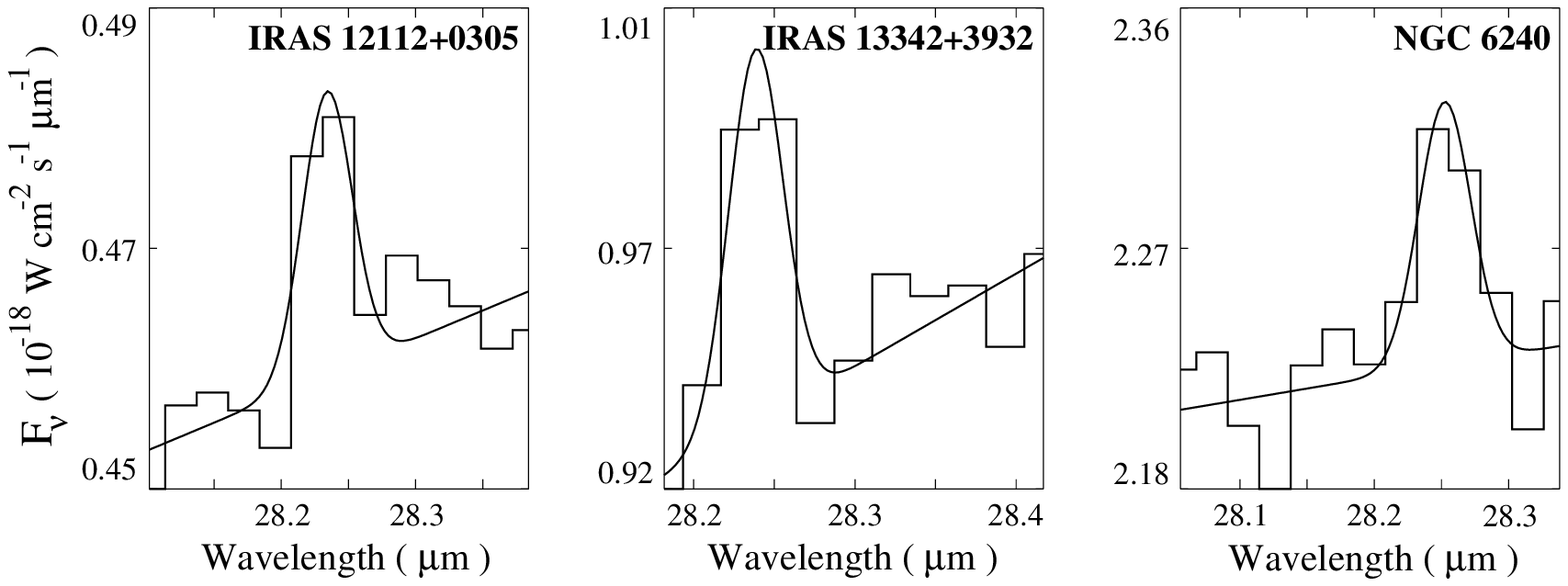}
\caption{Observed (histogram) and fitted (line) H$_{\rm
2}$ S(0) 28.22 $\mu$m lines detected with the IRS in the ULIRGs
IRAS~12112+0305, IRAS~13342+3932, and NGC~6240. The spectra represent
averages of the two slit positions. The vertical axis shows flux
density in units of 10$^{-18}$ W cm$^{-2}$ s$^{-1}$ $\mu$m$^{-1}$, while
the horizontal axis shows the rest wavelength in microns.}
\end{figure}

%FIGURE 3 Molecular Hydrogen s7

\begin{figure}
\centering
\label{fig:graphics:b}
\includegraphics[angle=0,scale=1]{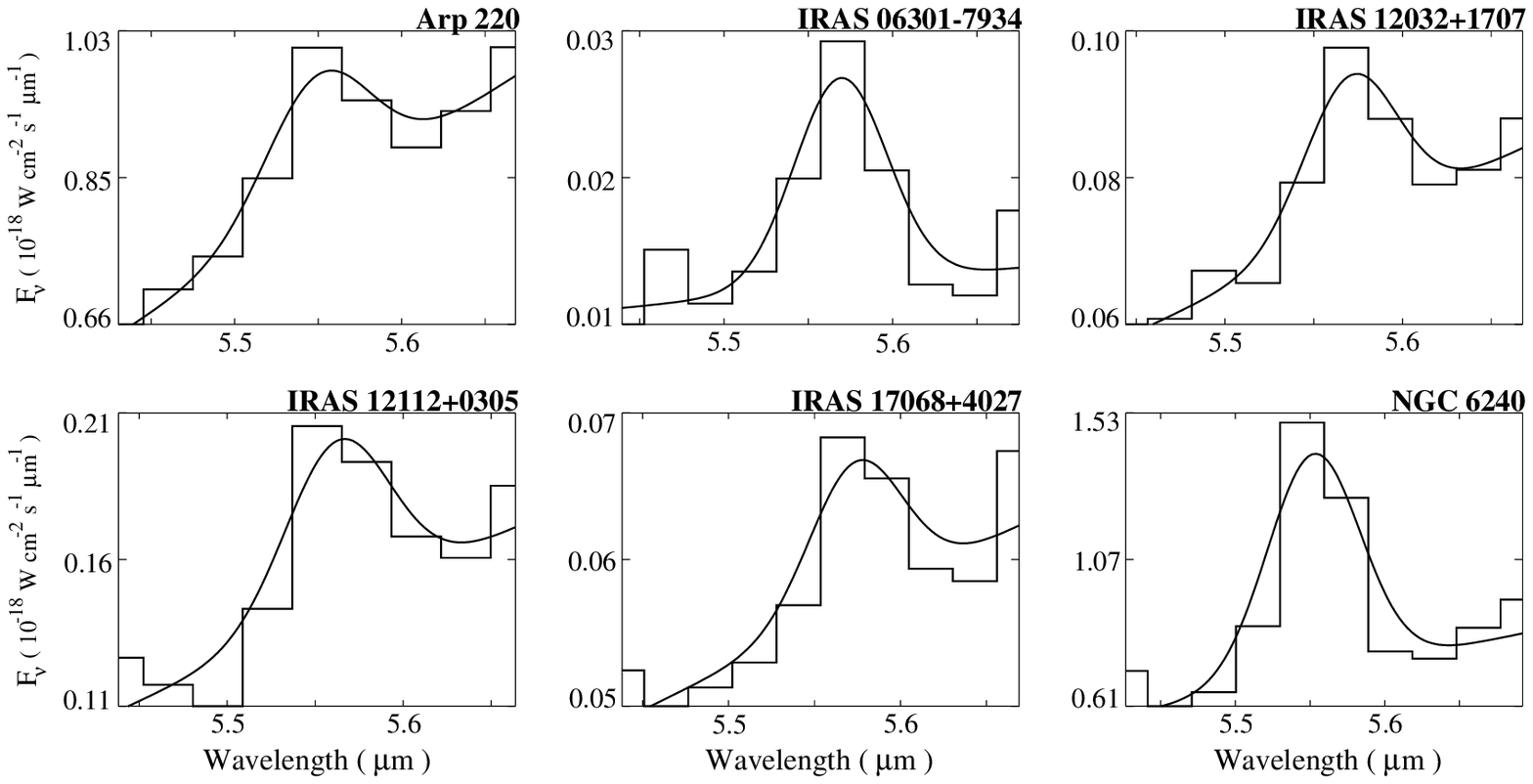}
\caption{Observed (histogram) and fitted (line) H$_{\rm 2}$
S(7) 5.51 $\mu$m emission lines detected with the IRS in the ULIRGs
Arp~220, IRAS~06301-79304, IRAS~12032+1707, IRAS~12112+0305,
IRAS~17068+4027, and NGC~6240. The spectra represent averages of
both slit positions. The vertical axis shows flux density in
units of 10$^{-18}$ W cm$^{-2}$ $\mu$m$^{-1}$.
Rest wavelength in microns is shown along the horizontal axis.}
\end{figure}

\clearpage

%FIGURE 4 single Excitations Diagram

\begin{figure}
\centering
\includegraphics[angle=0,scale=1]{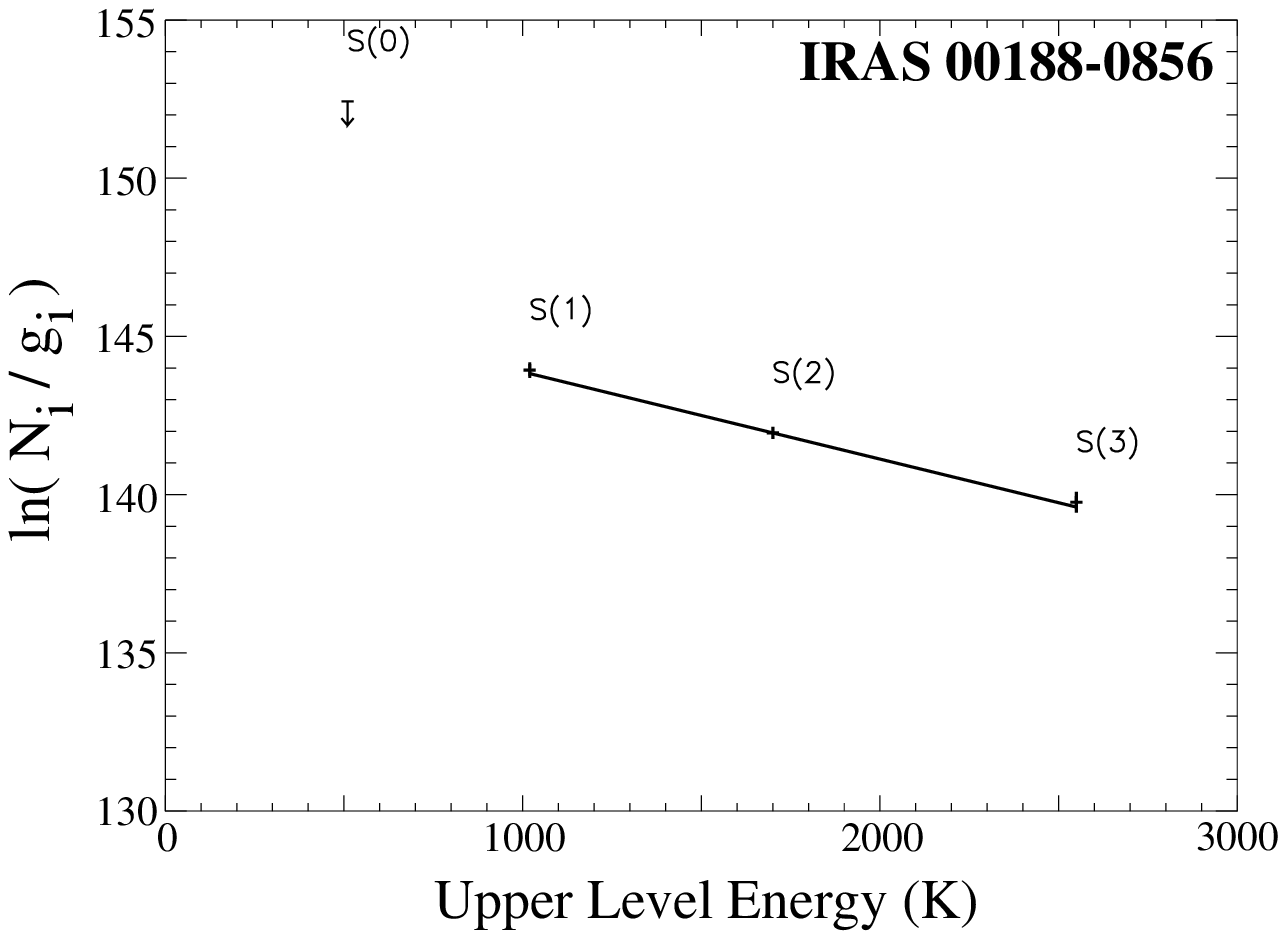}
\caption{Molecular hydrogen excitation diagram for IRAS~00188-0856,
which is typical for our ULIRG sample. The solid line shows a fit to
the S(1), S(2), and S(3) measurements, which is consistent with a
single temperature component at T$_{\rm ex}$ = 358 $\pm$ 25 K, with
a mass of 4.1 $\times$ 10$^{7}$ M$_{\odot}$. The arrow
shows the 3~$\sigma$ upper-limit for the S(0) transition.}
\end{figure}

%FIGURE 5 mrk273

\clearpage
\begin{figure}
\centering
%\label{fig:graphics:b}
\includegraphics[angle=0,scale=1]{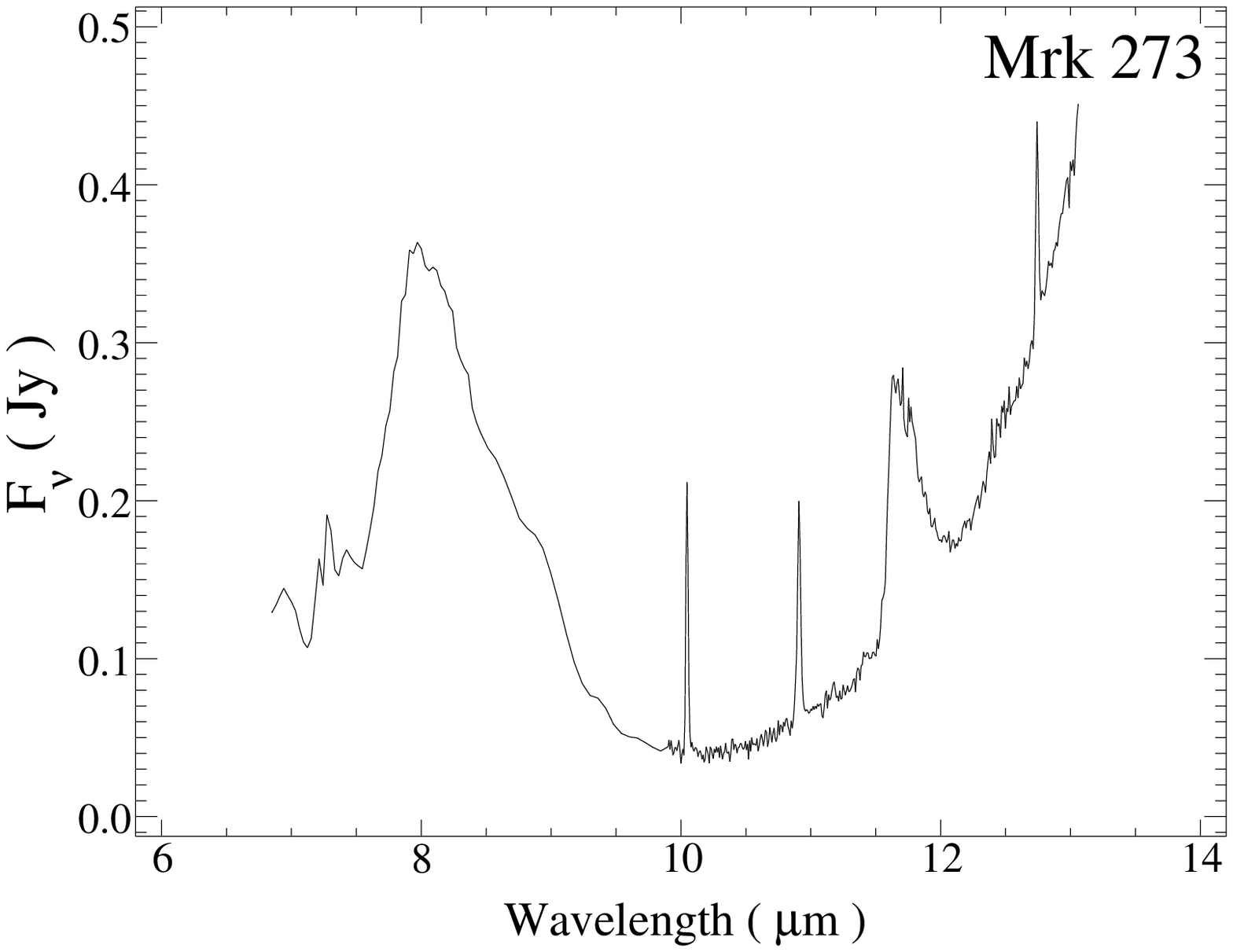}
\caption{ The combined IRS-LORES (6.5-9.8 $\mu$m) and IRS-HIRES
  (9.8-13.1 $\mu$m) spectrum for Mrk~273. The x-axis is the observed
  wavelength. A deep 9.7 $\mu$m silicate absorption trough is apparent,
  indicating heavy extinction (A$_{\rm V}$ $\sim$ 20-40) towards the
  active nucleus. On the other hand, the ratio of the S(3) 9.67 $\mu$m
  to S(2) 12.28 $\mu$m H$_{\rm 2}$ lines infers much lower levels of
  extinction, implying that little of the absorbing material lies in a
  screen between the warm H$_{\rm 2}$ and us.}
\end{figure}

\clearpage

%FIGURE 6 s1s3
\begin{figure}
\centering
\includegraphics[angle=0,scale=1]{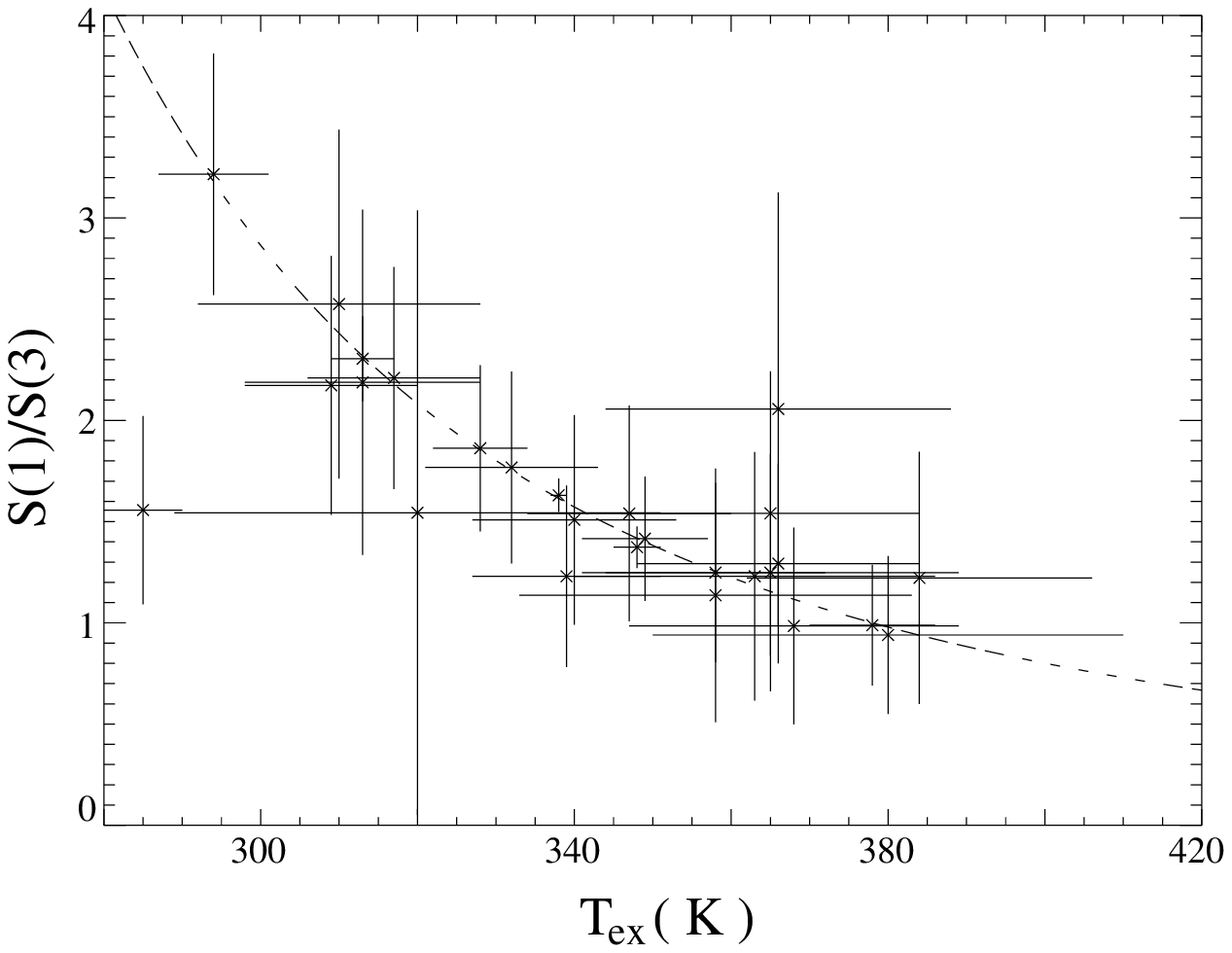}
\caption{ Molecular hydrogen S(1)/S(3) line ratios as a function of the
derived temperature (T$_{\rm ex}$ for twenty-seven ULIRGs (excluding Arp~220
and IRAS~12112+0305) with at least three J $\le$ 4 HIRES line 
measurements. The intrinsic (unreddened) ratio is shown as a dashed
line. Extinction towards the warm H$_{\rm 2}$ component is minimal in
most of the sources.}
\end{figure}

%FIGURE 7 s1s2
\begin{figure}
\centering
\includegraphics[angle=0,scale=1]{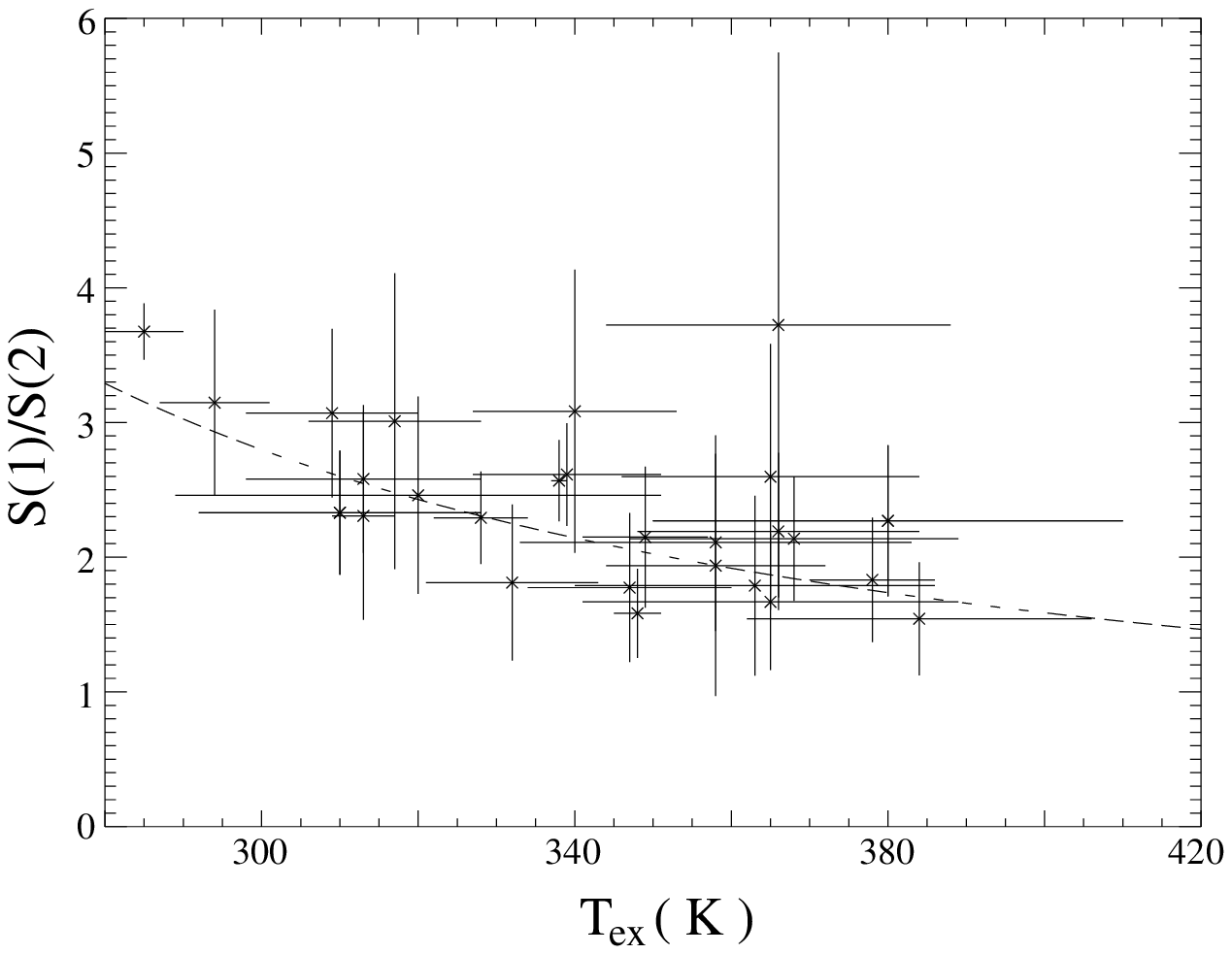}
\caption{ Molecular hydrogen S(1)/S(2) line ratios as a function of
  the derived excitation temperature (T$_{\rm ex}$) for the
  twenty-seven ULIRGs in Figure 8. The dashed line shows the expected
  trend for an ortho-to-para ratio of three and no extinction, which
  is consistent with the data.}
\end{figure}

\clearpage

% FIGURE 8 NGC 6240
\clearpage
\begin{figure}
\centering
\includegraphics[angle=0,scale=1.5]{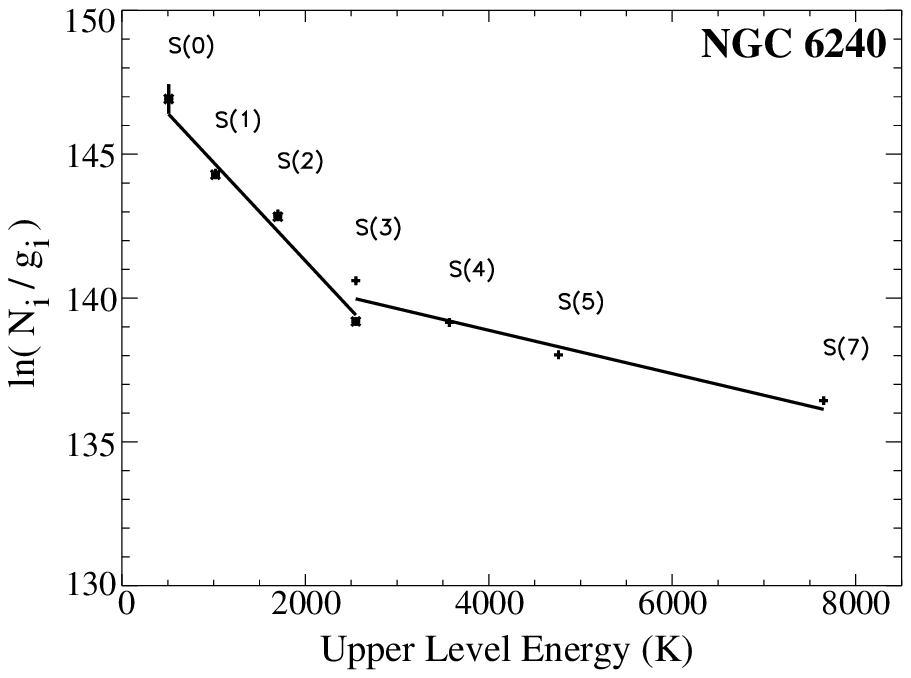}
\caption{Molecular hydrogen excitation diagram for NGC~6240. 
The S(3) through S(7) values can be fit by with a hot (T$_{\rm ex}$ = 
1327 $\pm$ 36 K) H$_{\rm 2}$ component. A second cooler (T$_{\rm ex}$ = 
292 $\pm$ 6 K) gas component is found after subtracting this component 
from the S(0) to S(3) values (re-plotted as asterisks) prior to modeling. }
\end{figure}

\clearpage

%FIGURE 9a
\begin{figure}
\centering
\includegraphics[angle=0,scale=.7]{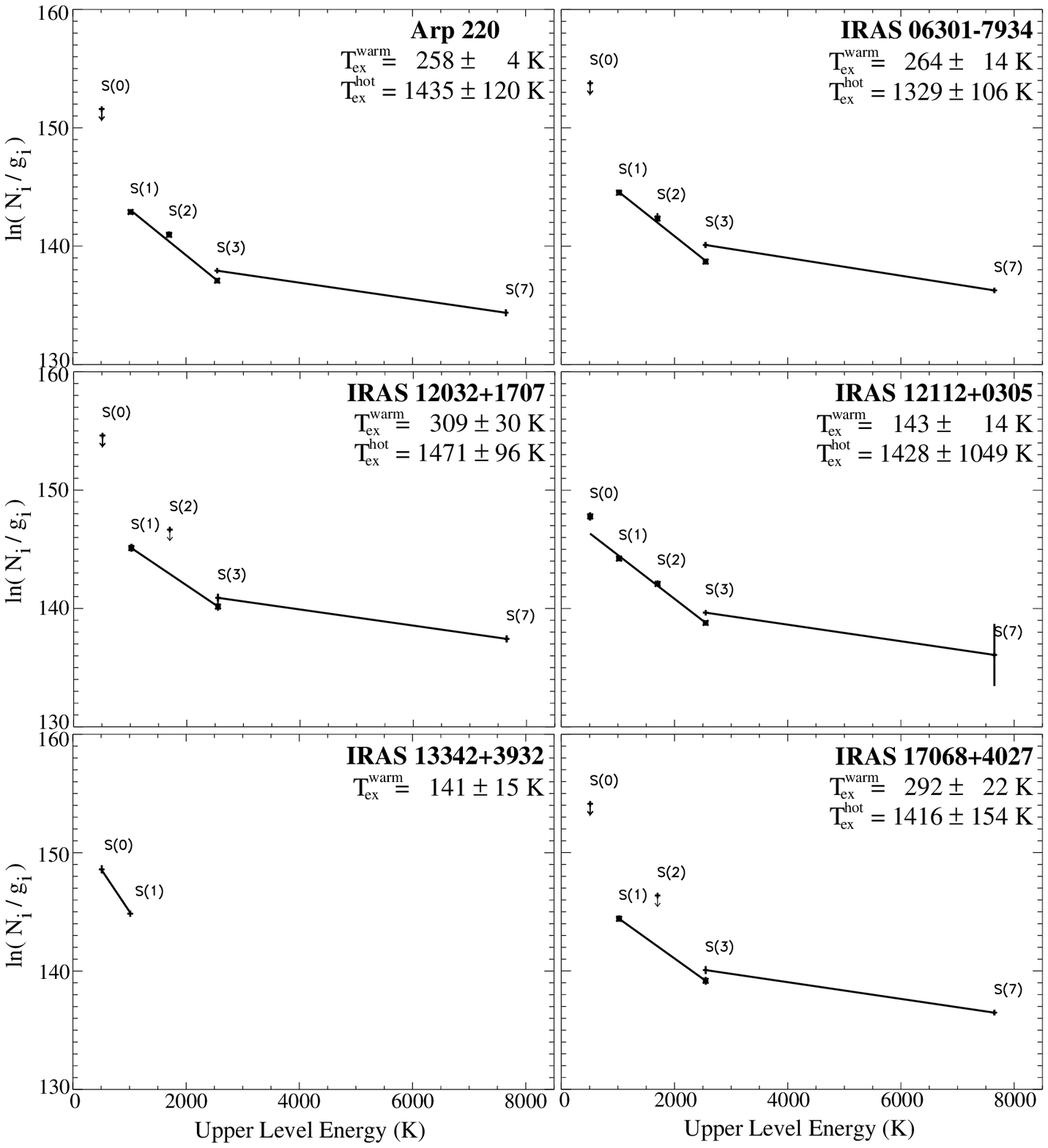}
\caption{Excitation diagrams for the six ULIRGs with either S(0) and/or
  S(7) line detections. In five sources, the S(7) and
  S(3) values are fit by a single hot H$_{\rm 2}$ component, which is
  subtracted from the S(0) through S(3) values (or limits) to
  characterize molecular hydrogen gas at a cooler temperature. 
  The derived warm (T$^{\rm warm}_{\rm ex}$) and hot (T$^{\rm hot}_{\rm ex}$)
  excitation temperatures are given in the upper-right of each panel. }
\end{figure}
\clearpage

%FIGURE 10 15 h2 irascolor

\begin{figure}
\centering
\includegraphics[angle=0,scale=1.]{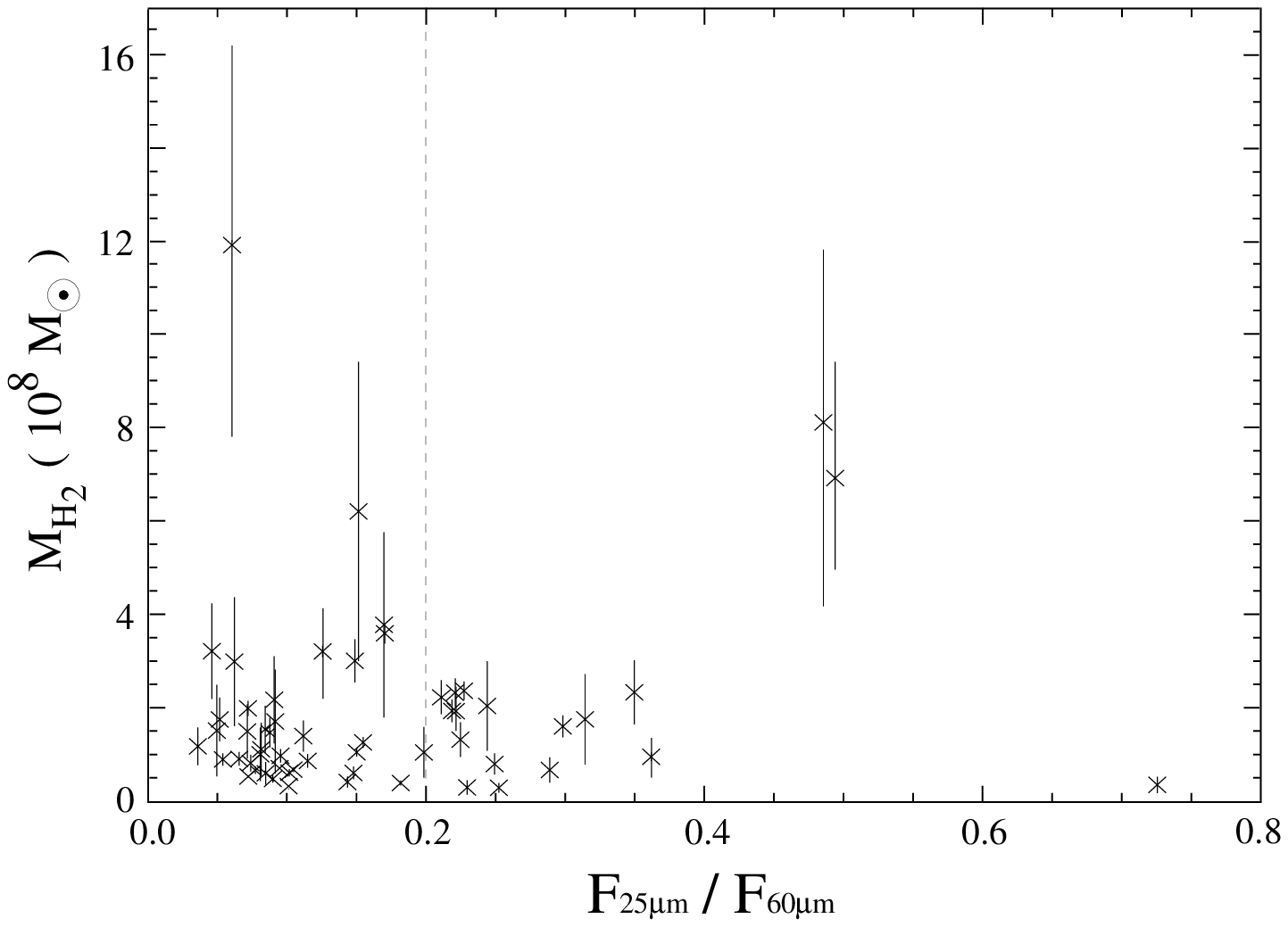}
\caption{The derived warm H$_{\rm 2}$ masses for the ULIRG sample
as a function of their IRAS 25 $\mu$m to 60 $\mu$m flux density ratios. 
The vertical line (F$_{\rm 25 \mu m}$/F$_{\rm 60 \mu m}$ = 0.2) is 
thought to separate sources dominated by starbursts 
(F$_{\rm 25 \mu m}$/F$_{\rm 60 \mu m}$ $<$ 0.2) from those powered
primarily by an enshrouded active nucleus 
(F$_{\rm 25 \mu m}$/F$_{\rm 60 \mu m}$ $>$ 0.2). 
No obvious relation between warm H$_{\rm 2}$ mass and 
F$_{\rm 25 \mu m}$/F$_{\rm 60 \mu m}$ is is present.}
\end{figure}

\clearpage

%FIGURE 11 16 h2 iras60

\begin{figure}
\centering
\includegraphics[angle=0,scale=1.5]{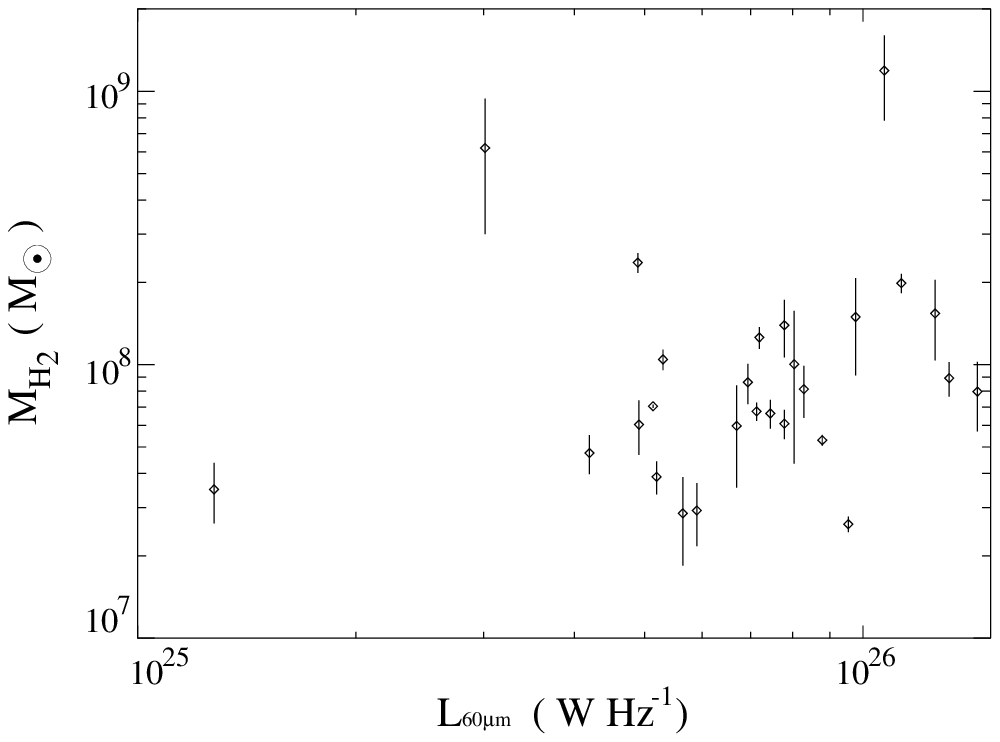}
\caption{Warm H$_{\rm 2}$ masses as a function of the IRAS
derived 60 $\mu$m specific luminosity (L$_{\rm 60 \mu m}$, in 
W Hz$^{-1}$) for the twenty-seven ULIRGs with z $<$ 0.1. There
is no clear tendency for more 60 $\mu$m luminous ULIRGs to possess
larger M$_{\rm H_2}$.}
\end{figure}

\clearpage

%FIGURE 12 17 warm gas fraction irascolor
\begin{figure}
\centering
\includegraphics[angle=0,scale=1.5]{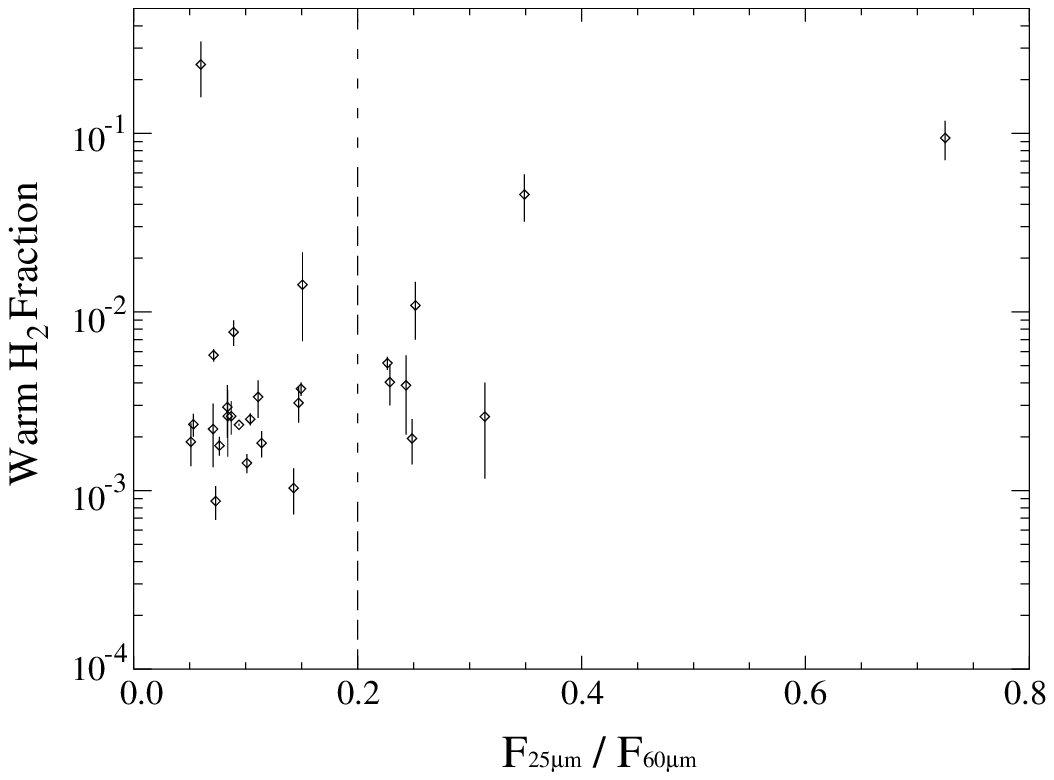}
\caption{The ratio of the warm H$_{\rm 2}$ mass to the cold H$_{\rm 2}$
mass from $^{12}$CO observations as a fraction of the IRAS 25 $\mu$m
to 60 $\mu$m flux density ratio. The dashed line separates ``cool'' and
``warm'' ULIRGs, i.e., those that appear to be powered primarily by
obscured starbursts or active nuclei, respectively.}
\end{figure}

\clearpage

\end{document}